\pdfoutput=1
\documentclass[12pt,a4paper]{article}
\usepackage{a4wide}
\usepackage{subcaption}
\usepackage[english]{babel}
\usepackage{amsmath}   
\usepackage{graphicx}  
\usepackage{dcolumn}   
\usepackage{bm}        
\usepackage{amssymb}   
\usepackage{epsfig}
\usepackage{epstopdf}
\usepackage{mdframed}
\usepackage{wasysym}
\usepackage[toc]{appendix}
\usepackage{color,soul} 
\usepackage{datetime}
\usepackage{physics}
\usepackage{enumerate}
\usepackage{tablefootnote}
\usepackage{multirow}

\usepackage{footmisc}

\usepackage{hyperref}

\newcommand{\be}{\begin{equation}}
\newcommand{\ee}{\end{equation}}
\newcommand{\bea}{\begin{eqnarray}}
\newcommand{\eea}{\end{eqnarray}}
\newcommand{\bean}{\begin{eqnarray*}}
\newcommand{\eean}{\end{eqnarray*}}

\newcommand{\mm}{\mathtt{m}}
\newcommand{\sm}{\mathtt{s}}

 \def\indirect{\emph{indirect} }
 \def\Indirect{\emph{Indirect} }
 \def\direct{\emph{direct BPS} }
  \def\Direct{\emph{Direct BPS} }
  
\begin{document}

\begin{titlepage}

\numberwithin{equation}{section}
\begin{flushright}
\small

\normalsize
\end{flushright}
\vspace{0.8 cm}

\begin{center}

\mbox{{\LARGE \textbf{On Supersymmetric Multipole Ratios}}}

\medskip

\vspace{1.2 cm} {\large Bogdan Ganchev$^1$, Daniel R. Mayerson$^{2}$}\\

\vspace{1cm} {$^1$ Universit\'e Paris-Saclay, CNRS, CEA,\\ Institut de Physique Th\'eorique, Orme des Merisiers\\ 91191, Gif-sur-Yvette CEDEX, France.}

\vspace{1cm} {$^2$ Instituut voor Theoretische Fysica, KU Leuven,\\ Celestijnenlaan 200D, B-3001 Leuven, Belgium}

\vspace{0.5cm}

\vspace{.5cm}
bogdan.ganchev @ ipht.fr, daniel.mayerson @ kuleuven.be

\vspace{2cm}

\textbf{Abstract}
\end{center}

Four-dimensional supersymmetric black holes are static and so have all vanishing multipoles. Nevertheless, it is possible to define finite multipole ratios for these black holes, by taking the ratio of (finite) multipoles of supersymmetric multicentered geometries and then taking the black hole scaling limit of the multipole ratios within these geometries.
An alternative way to calculate these multipole ratios is to deform the supersymmetric black hole slightly into a non-extremal, rotating black hole, calculate the multipole ratios of this altered black hole, and then take the supersymmetric limit of the ratios.
Bena and Mayerson observed that for a class of microstate geometries, these two a priori completely different methods give spectacular agreement for the resulting supersymmetric black hole multipole ratios. They conjectured that this agreement is due to the smallness of the entropy parameter for these black holes.
We correct this conjecture and give strong evidence supporting a more refined conjecture, which is that the agreement of multipole ratios as calculated with these two different methods is due to both the microstate geometry and its corresponding black hole having a property we call ``large dipole'', which can be interpreted as their center of mass being far away from its apparent center.

\end{titlepage}

\newpage

\setcounter{tocdepth}{2}
\tableofcontents

\newpage
\section{Introduction and Summary}\label{sec:intro}

When observing fields sourced by extended objects from afar, multipole expansions are invaluable tools that encode the structure of the source. For gravitating objects, one obtains gravitational multipoles as coefficients of an asymptotic expansion of their metric. These must be defined in a coordinate-invariant way --- as pioneered by Geroch \cite{Geroch:1970cd} and Hansen \cite{Hansen:1974zz} for stationary, vacuum spacetimes; later, Thorne \cite{Thorne:1980ru} defined an alternative way to define these multipoles, which was later shown to be equivalent to the Geroch-Hansen formalism in vacuum \cite{Gursel:equiv} and recently for non-vacuum spacetimes as well \cite{Mayerson:2022ekj}.

Gravitational multipoles come in two families for stationary solutions in four dimensions: the mass multipoles $M_\ell$ --- which can loosely be thought of as the coefficients in the $1/r$ expansion of $g_{tt}$, and the current multipoles $S_\ell$ --- which are related to the $1/r$ expansion of $g_{t\phi}$. (Note that we assume axisymmetry for simplicity). These multipoles characterize the metric; in vacuum, it can even be shown that a metric is completely determined by its multipole structure \cite{beig1980proof,kundu1981multipole,kundu1981analyticity}. For example, Kerr has multipoles $M_{2n} = M(-a^2)^n$ and $S_{2n+1} = Ma(-a^2)^n$, with all other multipoles vanishing.

The multipole moments of a gravitating object also show up in gravitational wave experiments --- e.g. deviations of a black hole's quadrupole from the Kerr  prediction $M_2 = -Ma^2 = -J^2 /M$ would indicate beyond-GR physics; such deviations have already been searched for in current gravitational wave observations \cite{Krishnendu:2017shb,LIGOScientific:2021sio}, and predictions have been made for the precision at which near-future experiments will be able to measure such deviations \cite{Krishnendu:2018nqa,Barack:2006pq,Gair:2017ynp,Fransen:2022jtw,Cano:2022wwo}.

The microstate geometry program suggests that black holes can be replaced by horizonless, smooth geometries. These essentially replace the horizon by quantum, stringy structure that is stable but nevertheless extends to (and a bit beyond) the horizon scale, without collapsing under its own gravitational force. (For reviews on microstate geometries and the parent fuzzball program, see e.g. \cite{Mathur:2005zp,Bena:2007kg,Bena:2013dka,Warner:2019jll,Shigemori:2020yuo}; and \cite{Mayerson:2020tpn,Mayerson:2022yoc} for an overview of its applicability in gravitational observations.) In the context of gravitational multipoles, it is interesting to wonder how the presence of near-horizon structure deforms the multipoles away from the black hole value. This was studied in \cite{Bena:2020see,Bena:2020uup,Bianchi:2020bxa,Bianchi:2020miz,Bah:2021jno}; the rough answer being that the multipoles of microstate geometries will approximate the black hole multipoles very well, up to corrections that depend precisely on the scale above the horizon at which the microstate geometry's structure becomes apparent \cite{Bah:2021jno}. A consequence is that a microstate geometry's multipoles will limit towards the black hole's multipoles in the scaling limit - the point at which the microstate geometry forms an infinite redshift throat and thus becomes the black hole itself.

Another quantity involving multipoles was introduced in \cite{Bena:2020see,Bena:2020uup}: multipole \emph{ratios}. These can be considered in microstate geometries, and one can then look at their behaviour in the scaling limit. If we are working with supersymmetric microstate geometries, their scaling limit is a static supersymmetric black hole whose multipoles (except the mass $M_0$) all vanish. However, even though the individual multipoles of the microstate geometries vanish in the scaling limit, the \emph{multipole ratios do not vanish}; they tend to a finite value in general. This suggests that these multipole ratios can be considered as intrinsic properties of the black hole itself; calculating these ratios in this way was called the \direct method.

Interestingly, there is a second, \indirect method to obtain finite multipole ratios for the same static, supersymmetric black holes. One can heat them up slightly into non-extremal, rotating black holes. Such non-extremal black holes have all multipoles non-vanishing, so the multipole ratios can be constructed; then, the supersymmetric extremal, non-rotating limit can again be taken of these multipole ratios.

A priori, the two methods to calculate multipole ratios for the supersymmetric black hole are entirely unrelated. It was then a great surprise in \cite{Bena:2020see,Bena:2020uup} that, for certain black holes and microstate geometries, \emph{these methods gave almost exactly the same results for the ratios}. This matching of multipole ratios was unexpected and unexplained, although it correlated with the black hole's charges having a small entropy parameter $\mathcal{H}\ll 1$; the natural conjecture put forward in \cite{Bena:2020see,Bena:2020uup} was then that a small entropy parameter was equivalent to the two multipole ratio methods agreeing in their calculation.

In this paper, we investigate further the necessary and sufficient conditions for the \direct and \indirect methods to give agreeing multipole ratios for a supersymmetric black hole. We show that the entropy parameter conjecture of \cite{Bena:2020see,Bena:2020uup} is not sufficient, and we give strong evidence for the correct necessary and sufficient conditions that both the microstate geometry (in the \direct method) and its corresponding supersymmetric black hole must satisfy. This then finally sheds some light on the mystery of why and when these completely different methods to calculate multipole ratios for supersymmetric black holes can agree.


\subsection{Summary}

We introduce the notion of having a ``large dipole'' for supersymmetric black holes as well as supersymmetric multicentered microstate geometries; this notion is the key feature of black holes and microstate geometries that relates their multipole ratios.

For supersymmetric black holes, we show that a ``large dipole'' black hole is synonymous with a ``small entropy parameter'' $\mathcal{H}$ black hole, i.e. black holes near the cosmic censorship bound have a large dipole $D$
\be \label{eq:DHequiv}
\frac{D}{M} \gg 1 \quad \Leftrightarrow \quad \mathcal{H} \ll 1.\ee
See Section \ref{sec:genBH} and for the definitions of the relevant quantities involved.

A ``large dipole'' microstate geometry is a geometry whose center positions $z_i=z_0+\tilde z_i$ in an ACMC frame satisfy $ z_0\gg |\tilde z_i|$. 
We give evidence that it is necessary and sufficient for a scaling microstate geometry and its corresponding supersymmetric black hole to satisfy this ``large dipole'' condition in order for the multipole ratios calculated using the \indirect method to agree well with the multipole ratios calculated in the \direct method:
\be \begin{aligned} \label{eq:conjTotal} \left(\frac{|D|}{M_\text{BH}}\gg 1\right) \text{ and } \left(\frac{z_0}{|\tilde z_i|}\gg 1\right) \text{ and } \left(\frac{J_\text{BH}}{aD} = -\frac{J_{\text{MG}}}{M_\text{BH}z_0}\right)\\
 \quad \Leftrightarrow \text{direct/indirect multipole ratios match}.
 \end{aligned}\ee
 This is derived in Sections \ref{sec:forward} \& \ref{sec:3centres}.
The additional condition involving the microstate geometry and black hole angular momenta $J_\text{MG},J_\text{BH}$ is necessary for the mixed (mass and current) multipole ratios to match as well. We conjecture (in Section \ref{sec:moreconj}) that this additional condition will in fact always be a consequence of the other two (large dipole) conditions.

It was conjectured in \cite{Bena:2020see,Bena:2020uup} that a small entropy parameter would be equivalent to \direct and \indirect methods giving matching answers for multipole ratios. However, we give an explicit counterexample (in Section \ref{sec:excounter}) that shows that $\mathcal{H}\ll 1$ (so that $D/M\gg1$) is not sufficient for the matching of the multipole ratios: the scaling microstate geometry used to calculate the \direct ratios must \emph{also} have a large dipole for the ratios to match. In this way, we have corrected and refined the conjecture of \cite{Bena:2020see,Bena:2020uup} for the matching of multipole ratios.

\medskip

The rest of this paper is devoted to showing the above relations. In Section \ref{sec:BG}, we briefly review the definitions of gravitational multipoles, as well as the relevant parameters and multipoles of the most general eight-charge black hole in four dimensions, our setup for microstate geometries, and the precise definitions of the \direct and \indirect methods of calculating multipole ratios for supersymmetric black holes. Section \ref{sec:BHtwocenter} is a short section devoted to displaying a curious analogy between a general black hole and a two-center microstate geometry. Then, in Section \ref{sec:matchingproofs} we come to the main results of our paper, where we give strong evidence for the equivalence between the property of having a ``large dipole'' for the black hole and microstate geometry, and the matching of the multipole ratios. Section \ref{sec:examples} gives a number of explicit examples supporting our arguments. Finally, we give a brief discussion and further (refined) conjectures in Section \ref{sec:moreconj}.

\section{Background}\label{sec:BG}

In this section, we review some background information on gravitational multipoles, general four-dimensional black holes, and supersymmetric multicentered microstate geometries. We focus on reviewing (only) the crucial features of each concept or object that we will need and use in this paper. At the end of this section, we contrast the two notions of multipole ratios that is the main focus of our analysis.

\subsection{Gravitational multipoles}
Asymptotically flat, four-dimensional, stationary metrics have two families of coordinate-independent multipole moments; these were first described by Geroch \cite{Geroch:1970cd} and Hansen \cite{Hansen:1974zz} for vacuum solutions, and later generalized in various ways \cite{Sotiriou:2004ud,Pappas:2014gca,Fodor:2020fnq,Cano:2022wwo,Mayerson:2022ekj}. Thorne later introduced the notion of asymptotically Cartesian and mass-centered (ACMC) coordinates, in which a stationary, axisymmetric metric\footnote{We will only discuss axisymmetric metrics for simplicity; the non-axisymmetric generalization is straightforward and given in e.g. appendix A of \cite{Bena:2020uup}. Note that we are assuming ACMC-$\infty$ coordinates; Thorne has a more general formulation where the coordinates can be only ACMC to a certain order $N$ \cite{Thorne:1980ru}.} has the particular asymptotic expansion in powers of $1/r$:
\begin{align}
\label{eq:gttmultipoles}  g_{tt}  &= -1 + \frac{2M_\text{BH}}{r}+ \sum_{\ell\geq 1}^{\infty}  \frac{2}{r^{\ell+1}} \left( \tilde M_\ell P_\ell + \sum_{\ell'<\ell} c^{(tt)}_{\ell \ell'} P_{\ell'} \right), \\
\label{eq:gtphimultipoles}  g_{t\phi} &= -2r \sin^2\theta\left[ \sum_{\ell\geq 1}^{\infty} \frac{1}{r^{\ell+1}} \left( \frac{\tilde S_\ell}{\ell} P'_\ell +\sum_{\ell'<\ell} c_{\ell \ell'}^{(t\phi)}  P'_{\ell'}\right) \right],\\
 \label{eq:gspacespacemultipoles}  g_{rr} &= 1 + \sum_{\ell\geq 0}^\infty\frac{1}{r^{\ell+1}} \sum_{\ell'\leq \ell} c_{\ell \ell'}^{(rr)} P_{\ell'}, & 
   g_{\theta\theta} &= r^2\left[ 1 + \sum_{\ell\geq 0}^\infty\frac{1}{r^{\ell+1}} \sum_{\ell'\leq \ell} c_{\ell \ell'}^{(\theta\theta)} P_{\ell'}\right],\\ 
g_{\phi\phi} &= r^2\sin^2\theta\left[ 1 + \sum_{\ell\geq 0}^\infty\frac{1}{r^{\ell+1}} \sum_{\ell'\leq \ell} c_{\ell \ell'}^{(\phi\phi)} P_{\ell'} \right], &
\nonumber g_{r\theta} &= (-r\sin\theta) \left[ \sum_{\ell\geq 0}^\infty\frac{1}{r^{\ell+1}} \sum_{\ell'\leq \ell} c_{\ell \ell'}^{(r\theta)} P'_{\ell'}\right],
\end{align}
where $P_\ell$ denotes a Legendre polynomial.
The coefficients $M_\ell$ are the \emph{mass multipoles} and the coefficients $S_\ell$ are the \emph{current multipoles} (or angular momentum multipoles); the mass and angular momentum are simply $M_\text{BH}=M_0$ and $J_\text{BH}=S_1$. Note that $S_0=0$, and that part of the definition of the ACMC (namely, being \emph{mass centered}) coordinate system is such that the mass dipole vanishes, $M_1=0$.

Kerr, with mass $M_\text{BH}$ and angular momentum $J_\text{BH}=a M_\text{BH}$, has a simple collection of multipole moments:
\be M_{2n} = M_\text{BH}(-a^2)^n, \qquad S_{2n+1} = aM_\text{BH}(-a^2)^n, \qquad M_{2n+1}=S_{2n}=0.\ee
It is interesting to note that Kerr-Newman has precisely the same multipole moments, i.e. they are independent of the charge of the Kerr-Newman black hole.

It will be useful to us to use a ``master function'' where both the mass and current multipoles are collected into a single complex, harmonic function:\footnote{This was introduced in \cite{Bianchi:2020bxa}, although it is not clear if this is the first occurrence of this idea.}
\be H \equiv \sum_{\ell=0}^\infty \frac{M_\ell + i S_\ell}{r^{\ell+1}}P_{\ell}(\cos\theta).\ee
For example, for Kerr this gives the very elegant and simple expression:
\be H_{\text{Kerr}} = \frac{M_\text{BH}}{\sqrt{x^2+y^2+(z-i\, a)^2}} ,\ee
where we use the standard relation between Cartesian and spherical, $(r,\theta,\varphi)$, coordinates.

\subsection{General eight-charge black holes}\label{sec:genBH}
We will consider the most general non-extremal black hole in asymptotically flat four dimensions (STU BH), described by Chow and Comp\`ere in \cite{Chow:2014cca}. The metric of these stationary black holes depends on a mass parameter $m$, a rotation parameter $a$, and eight charge parameters $\delta_I,\gamma_I$ ($I=0,1,2,3$). Such a black hole, with mass $M_\text{BH}$ and angular momentum $J_\text{BH}$, can have four electric charges $Q_I$ and four magnetic charges $P^I$; $M_\text{BH},J_\text{BH},Q_I,P^I$ are complicated functions of the ten parameters $m,a,\delta_I,\gamma_I$ --- see Appendix \ref{app:chargeparams} for the full expressions.

The multipoles of such a generic eight-charge black hole were derived in \cite{Bena:2020see,Bena:2020uup}:
\begin{align}
\nonumber  M_\ell &= -\frac{i}{2} \left( -\frac{a}{M_\text{BH}}\right)^\ell Z \bar{Z} (Z^{\ell-1}-\bar{Z}^{\ell-1}),\\
\label{eq:MSgenBH} S_\ell &= \frac{i}{2} \left( -\frac{a}{M_\text{BH}}\right)^{\ell-1} \frac{J_\text{BH}}{M_\text{BH}} (Z^\ell-\bar{Z}^\ell),\\
 \nonumber Z & \equiv D - iM_\text{BH}, & \bar{Z} = D + i M_\text{BH},
 \end{align}
 where the mass $M_\text{BH}$, angular momentum $J_\text{BH}$, and dipole parameter $D$ are again complicated functions of $m,a,\delta_I,\gamma_I$ (see Appendix \ref{app:chargeparams}). Remarkably, it is still possible to collect these multipoles in a single harmonic function $H$ with two (complex) poles, just as for Kerr:
 \begin{align} H_{\text{BH}} &= \sum_{\ell=0}^\infty \frac{M_\ell + i S_\ell}{r^{\ell+1}}P_{\ell}(\cos\theta) \\
 &= \frac12 \frac{M_\text{BH}- J_\text{BH}/a  + i D}{\sqrt{x^2+y^2+(z+Da/M_\text{BH} + i a)^2}} +\frac12 \frac{M_\text{BH}+ J_\text{BH}/a  - i D}{\sqrt{x^2+y^2+(z+Da/M_\text{BH} - i a)^2}}.
 \label{eq:HBH}
 \end{align}

The supersymmetric extremal limit of these black holes is given by a particular scaling of the parameters in which $m,a\rightarrow 0$ while the charges $Q_I,P^I$ are kept fixed (see Appendix \ref{app:chargeparams}). We will only be interested in supersymmetric black holes with charges:
\be P^1=P^2 = 0, \qquad P^0 = -P^3,\ee
and moreover we will assume that:
\be Q_0\ll Q_1,Q_2,Q_3.\ee
Note that in the supersymmetric limit, $a\rightarrow 0$ but e.g. $M_\text{BH},D$ and $J_\text{BH}/a$ stay finite.

For supersymmetric black holes, we can define the \emph{entropy parameter} $\mathcal{H}$ as:
\be \label{eq:Hdef} \mathcal{H} = \frac{\mathcal{Q}(Q_I,P^I)}{Q_0Q_1Q_2Q_3},\ee
where $\mathcal{Q}(Q_I,P^I)$ is the \emph{quartic invariant} for the black hole which gives (the square of) its entropy. When $P^I=0$, this quartic invariant reduces to $\mathcal{Q}(Q_I,P^I=0)=Q_0Q_1Q_2Q_3$, and so the entropy parameter $\mathcal{H}$ gives a dimensionless notion of how small the entropy of the black hole with charges $Q_I,P^I$ is compared to the black hole with equal electric charges $Q_I$ but zero magnetic charges $P_I$.\footnote{The five-dimensional interpretation of $\mathcal{H}$ is that it quantifies how close to the cosmic censorship bound the black hole is; for precisely $\mathcal{H}=0$, the corresponding black hole is on the bound and has exactly vanishing horizon area.} For the full, eight-charge expression of $\mathcal{Q}$, see e.g. \cite{Bena:2020uup}; when $P^1=P^2=0$ and $P^0=-P^3$, we have:
\be \mathcal{Q} = Q_0 Q_1Q_2Q_3 -\frac14 (Q_0+Q_3)^2 P_3^2.\ee

Supersymmetric black holes in four dimensions are static and so all their multipoles (except $M_0$) vanish. However, we can define \emph{multipole ratios} $\mathcal{M}$ for these black holes \cite{Bena:2020see,Bena:2020uup}, taken as a \emph{limit} of the multipole ratio of a general, non-extremal black hole with the same charges $Q_I,P^I$:
\be \mathcal{M}_{\text{indirect}} = \lim_{\text{BH}\rightarrow \text{SUSY}} \mathcal{M}(m,a,\delta_I,\gamma_I).\ee
The limit taken in this way is remarkably independent of the direction within the ten-dimensional parameter space $m,a,\delta_I,\gamma_I$ \cite{Bena:2020see,Bena:2020uup}, giving a well-defined quantity we can associate to the supersymmetric black hole. Calculating multipole ratios in this way for supersymmetric black holes, i.e. using a `detour' through a non-extremal version of the black hole with the same charges, is called the \indirect method of calculating multipole ratios for supersymmetric black holes.\footnote{The \indirect method can also be used to define multipole ratios for the Kerr black hole, see \cite{Bena:2020see,Bena:2020uup} and \cite{Cano:2022wwo} for an extension to higher-derivative theories.} See Appendix B of \cite{Bena:2020uup} for further explanation. For example, consider the following multipole ratio for generic (non-extremal) black holes:
\be \mathcal{M}_\text{ex} = \frac{M_2 M_2}{M_4 M_0} = - \frac{D^2 + M_\text{BH}^2}{3D^2 - M_\text{BH}^2},\ee
where we used (\ref{eq:MSgenBH}).
This last expression is manifestly finite and well-defined in the supersymmetric black hole limit (since $D,M_\text{BH}$ remain finite for supersymmetric black holes), and so gives a well-defined multipole ratio calculated for the supersymmetric black hole using the \indirect method.

\subsection{Supersymmetric multicentered microstate geometries}\label{sec:MGs}
A supersymmetric multicentered microstate geometry in four dimensions is completely determined by a collection of eight harmonic functions $(V,K^{\hat{I}},L_{\hat{I}},M)$, with $\hat{I}=1,2,3$:
\be \label{eq:MGharmfuncs}\begin{aligned}
 V &= 1 + \sum_{i=1}^n \frac{v_i}{r_i}, & K^{\hat{I}} &= \sum_{i=1}^n \frac{k^{\hat{I}}_i}{r_i},\\
 M &= \sum_{i=1}^n \frac{m_i}{r_i}, & L_{\hat{I}} &= 1 + \sum_{i=1}^n \frac{l^{\hat{I}}_i}{r_i},
\end{aligned}\ee
The constant terms are the moduli of the solution (which we have chosen here such that the solution is asymptotically flat in four dimensions), and the harmonic functions are further completely determined by the spatial location of their poles, $\vec{r}_i$, where $r_i$ is the distance to each pole, $r_i\equiv |\vec{r}-\vec{r}_i|$, as well as the charges at each pole. These poles are precisely the $n$ ``centers'' of the geometry, and in a five-dimensional uplift, there are topological ($S^2$) ``bubbles''  stretching between each pair of centers, giving the geometry a non-trivial topology. The charges of such a solution are:
\begin{align}
 Q_{\hat{I}} &= \sum_i l^{\hat{I}}_i, & Q_0 &= \sum_i v_i,\\
 P^{\hat{I}} &= \sum_i k^{\hat{I}}_i & P^0 &= -2 \sum_i m_i.
\end{align}
We will always work with microstates that have only two magnetic charges non-zero, e.g. $P^0 = -P^3$ and $P^1=P^2=0$. (Note that $\sum_I P^I = 0$ is always necessary from the bubble equations for our chosen moduli.) Other details of these geometries, such as their metric given in terms of these eight harmonic functions, are given in appendix \ref{app:MGs}.

The center charges and intercenter distances must satisfy complicated non-linear relations called \emph{bubble equations} to ensure the geometry is regular. Such a microstate geometry is entirely \emph{smooth} (i.e. without singularities) in a five-dimensional uplift if the $l_i^{\hat{I}},m_i$ charges satisfy:
\be \label{eq:MGsmoothness} l_i^1 = -\frac{k^2_ik^3_i}{v_i}, \qquad m_i = \frac12 \frac{k^1_ik^2_ik^3_i}{v_i^2},\ee
and cyclic permutations of $(1,2,3)$, for all centers $i$.


We will limit ourselves to work with smooth microstate geometries that are approximately axisymmetric, so that $\vec{r}_i=(\mathcal{O}(\tilde{\epsilon}),0,z_i)$, with $\tilde{\epsilon}\ll z_i$, and that admit a \emph{scaling} limit; this means there exists a continuous family of solutions with the same asymptotic charges\footnote{For exactly axisymmetric scaling solutions, this condition must be relaxed to be \emph{approximately} the same charges in the scaling limit.} where all the centers limit towards coinciding, $\lvert\vec{r}_i\rvert\rightarrow 0$. At the \emph{scaling point} ($\lvert\vec{r}_i\rvert=0$), the geometry becomes a supersymmetric black hole.

The multipoles of supersymmetric and axisymmetric microstate geometries are given by \cite{Bena:2020see,Bena:2020uup}:
\begin{align} \label{eq:MGmultipoles} M_\ell &= \sum_i \mm_i z_i^\ell, & S_\ell &= \sum_i \sm_i z_i^\ell,\\
 \mm_i & \equiv \frac14 \left( v_i + l^1_i + l^2_i + l^3_i\right), & \sm_i &\equiv \frac14\left( -2m_i + k^1_i + k^2_i + k^3_i\right), 
\end{align}
taken in a coordinate system where the origin is chosen such that $M_1 =0$. We can again summarize the multipoles in a single complex harmonic function as:
\be\label{eq:HMG} H_{\text{MG}} =  \sum_{\ell=0}^\infty \frac{M_\ell + i S_\ell}{r^{\ell+1}}P_{\ell}(\cos\theta) = \sum_i \frac{\mm_i + i \sm_i}{\sqrt{x^2+y^2+(z-z_i)^2}}.\ee
Note that this was already noticed explicitly in \cite{Bianchi:2020bxa,Bianchi:2020miz}.

We can again consider multipole ratios $\mathcal{M}$ in supersymmetric, axisymmetric microstate geometries; the scaling limit $z_i\rightarrow 0$ of these ratios is then again well-defined, even though the scaling point supersymmetric black hole is again static and has vanishing multipoles itself. We call this method of computing multipole ratios for supersymmetric black holes the \direct method.

\subsection{Multipole ratios calculated in two ways}\label{sec:2ways}
Even though supersymmetric black holes have all vanishing multipoles, we introduced two distinct methods for calculating their multipole \emph{ratios} $\mathcal{M}$ as a well-defined limit of other, related objects. Using the \indirect method, we calculate the multipole ratios for a non-extremal black hole with the same charges, before taking the supersymmetric extremal limit:
\be \mathcal{M}_{\text{indirect}} = \lim_{\text{BH}\rightarrow \text{SUSY}} \mathcal{M}_{\text{BH}}.\ee
The \emph{direct} method, on the other hand, uses the scaling limit of smooth, microstate geometries to find the ratios as a limit of the microstate geometry ratios:
\be \mathcal{M}_{\text{direct}} = \lim_{\text{(scaling)} z_i\rightarrow 0} \mathcal{M}_{\text{MG}}.\ee

These two methods of calculating or associating multipole ratios to supersymmetric black holes are \emph{a priori} entirely unrelated. It is then somewhat of a surprise that both of these methods give \emph{approximately the same result} for particular microstate geometries (but certainly not all), as was noticed in \cite{Bena:2020see,Bena:2020uup}. Evidence was given in \cite{Bena:2020see,Bena:2020uup} for the conjecture that the matching of the multipole ratios in \direct and \indirect methods was related to the smallness of the entropy parameter $\mathcal{H}$ of the supersymmetric black hole --- the smaller $\mathcal{H}$, the better the matching between the methods.

In this paper, we give strong evidence for a preciser and corrected version of this conjecture. We show what is the correct condition that is necessary and sufficient for the matching of the \direct and \indirect methods, and moreover quantify how much they are expected to (dis)agree.

\section{Black Holes as Two-Center Geometries}\label{sec:BHtwocenter}

The complex harmonic function $H_\text{MG}$ of (\ref{eq:HMG}) that determines the multipoles of a \emph{two-center} microstate geometry is:
\be H_{\text{MG}} =  \sum_{\ell=0}^\infty \frac{M_\ell + i S_\ell}{r^{\ell+1}}P_{\ell}(\cos\theta) =  \frac{\mm_1 + i \sm_1}{\sqrt{x^2+y^2+(z-z_1)^2}}+\frac{\mm_2 + i \sm_2}{\sqrt{x^2+y^2+(z-z_2)^2}}.\ee
Note that $\sm_1+\sm_2=0$ and the center positions must be chosen such that $\mm_1 z_1 =- \mm_2 z_2$ (to ensure that $M_1=0$).

This function $H_\text{MG}$ bears remarkable similarity to the harmonic function $H_\text{BH}$ given in (\ref{eq:HBH}) that determines the most general black hole multipoles. Functionally, to obtain $H_\text{BH}$ from $H_\text{MG}$, we must identify:
\be \label{eq:m12MGBH} \mm_{1,2} = M_\text{BH} \pm \frac{J_\text{BH}}{a}, \qquad \sm_{1,2} = \pm D,\ee
together with the center position identification:
\be \label{eq:z12MGBH} z_{1,2} = -a\frac{D}{M_\text{BH}} \left( 1 \mp i \left(\frac{D}{M_\text{BH}}\right)^{-1} \right), \ee
which amounts to a ``complex shift'' of the poles of the harmonic function $H_\text{MG}$.
At this point, we are simply rewriting the black hole as an effective, fictitious two-center geometry with complex center positions.

A general two-center geometry has 16 degrees of freedom, corresponding with choosing the eight center charges $(v_i, k^{\hat{I}}_i,l^{\hat{I}}_i, m_i)$ for each center. With 8 fixed charges $Q^I, P_I$, there are in principle still 8 degrees of freedom remaining. However, a \emph{smooth} two-center geometry must satisfy (\ref{eq:MGsmoothness}) at each center, which gives an additional 8 constraints fixing the center charges. In other words, a smooth two-center geometry is \emph{completely determined} by its charges $Q^I, P_I$.

If we could view the static, supersymmetric eight-charge black hole as a limit $J_\text{BH},a\rightarrow 0$ of the fictitious two-center \emph{smooth} geometry given in (\ref{eq:m12MGBH}) and (\ref{eq:z12MGBH}), then we can easily find the black hole parameters $D/M_\text{BH}$ and $J_\text{BH}/(aD)$ in terms of the charges on the two smooth centers. In particular, if we parametrize the magnetic charge $P_0$ as:
\be \label{eq:P0parameps} P_0 = 2 \frac{\sqrt{Q_0 Q_1 Q_2 Q_3}}{Q_0+Q_1}(1-\epsilon),\ee
which is chosen such that for small $\epsilon$, the entropy parameter (\ref{eq:Hdef}) is simply:
\be \label{eq:Hparameps} \mathcal{H} = 2\epsilon + \mathcal{O}(\epsilon^2),\ee
then we obtain:
\begin{align} \label{eq:DoverM} \left|\frac{D}{M_\text{BH}}\right|& = \frac{ |\mm_1-\mm_2|}{M_\text{BH}} =   \left| \frac{1}{\sqrt{2}}\frac{(Q_0 - Q_1)(Q_0 + Q_1 -Q_2-Q_3)}{(Q_0+Q_1)(Q_0+Q_1+Q_2+Q_3)} \frac{1}{\sqrt{\epsilon}}\right| + \mathcal{O}(\epsilon^{1/2}),\\
 \label{eq:JoveraD}  \frac{J_\text{BH}}{aD} &= -2\frac{\sm_1}{\mm_1-\mm_2} \approx  \left( \left( \frac{Q_0 Q_2}{Q_1 Q_3}\right)^{1/2}+\left(\frac{Q_0 Q_3}{Q_1 Q_2}\right)^{1/2}-\left(\frac{Q_0 Q_1}{Q_2 Q_3}\right)^{1/2}\right)^{-1} + \mathcal{O}\left(\frac{Q_0}{Q_{\hat{I}}}\right).
\end{align}
where in (\ref{eq:DoverM}) we used $\epsilon\ll 1$, and in (\ref{eq:JoveraD}) we used $Q_0\ll Q_{\hat{I}}$ (but did not need $\epsilon\ll 1$).

For a supersymmetric black hole, the parameters $D$ and $J_\text{BH}/a$ must be completely determined by the black hole charges $Q_I, P^I$. However, a priori there is no reason why (\ref{eq:DoverM}) and (\ref{eq:JoveraD}) should be the correct ones for a supersymmetric black hole --- after all, we obtained these expressions by invoking a fictitious, complex, smooth two-center microstate geometry! It is then remarkable that (\ref{eq:DoverM}) and (\ref{eq:JoveraD}) \emph{are indeed the correct expressions for a supersymmetric black hole}. This is very hard to check analytically --- both the charges $Q_I,P^I$ and the quantities $D, J_\text{BH}/a$ can be expressed in terms of the black hole parameters $\delta_I,\gamma_I$ (see appendix \ref{app:chargeparams}), but inverting these dependencies to find e.g. $D(Q_I,P^I)$ seems intractable; nevertheless, we have checked (\ref{eq:DoverM}) and (\ref{eq:JoveraD}) numerically over a wide range of supersymmetric black hole parameters and are confident of its correctness.

\section{Large Dipole Geometries and Matching Multipole Ratios}\label{sec:matchingproofs}

In this section, we investigate the necessary and sufficient conditions for the \direct and \indirect methods to give matching multipole ratios for supersymmetric black holes. We first introduce the notion of ``large dipole'' black holes and microstate geometries in Section \ref{sec:largeDipoleMS}; then we prove that the condition of large dipole is sufficient for the matching of ratios in Section \ref{sec:forward}. Finally, in Section \ref{sec:3centres}, we provide strong evidence that the condition of large dipoles is also necessary for this matching.

\subsection{Large dipole geometries}\label{sec:largeDipoleMS}
For black holes (non-extremal or supersymmetric), we define having a ``large dipole'' if $D$ is large, so that:
\be \label{eq:largedipoleBH} \frac{|D|}{M_\text{BH}}\gg 1.\ee
Note that from (\ref{eq:DoverM}) and (\ref{eq:Hparameps}), it follows that $|D|/M_\text{BH}\gg 1$ if and only if the entropy parameter is small, $\mathcal{H}\ll 1$.

For large dipole black holes, we can expand their multipoles (\ref{eq:MSgenBH}) to leading order in $D/M_\text{BH}\gg 1$:
\be \label{eq:multlargedipoleBH} \begin{aligned} M_\ell &= (1-\ell)M_\text{BH}\left( -a\frac{D}{M_\text{BH}}\right)^\ell + \mathcal{O}\left(\left(\frac{D}{M_\text{BH}}\right)^{\ell-1}\right),\\ S_\ell &= \ell J_{\text{BH}} \left( -a \frac{D}{M_\text{BH}}\right)^{\ell-1}+\mathcal{O}\left(\left(\frac{D}{M_\text{BH}}\right)^{\ell-2}\right).\end{aligned}\ee

For supersymmetric microstate geometries, the positions of the centers are fixed by the bubble equations and by demanding that the mass dipole vanishes, $M_1=0$. We can always reparametrize the positions of the centers as:
\be \label{eq:zitozitilde} z_i = z_0 + \tilde z_i .\ee
There is always some freedom in choosing $z_0$ in (\ref{eq:zitozitilde}) since we are writing the $n$ parameters $z_i$ in terms of $n+1$ parameters $z_0,\tilde z_i$.
We define that the microstate geometry has a ``large dipole'' if we can choose the $z_0,\tilde z_i$ such that:
\be \label{eq:largedipoleMG} \frac{z_0}{|\tilde z_i|}\gg 1,\ee
for all $i$. For a large dipole microstate geometry, using that $M_1=0$ implies $\sum_i \mm_i \tilde z_i = -M_\text{BH}z_0$, we can expand the multipoles to leading order in the $\tilde z_i$ parameters:
\be \label{eq:multlargedipoleMG} M_\ell = (1-\ell)   M_\text{BH} z_0^\ell + \mathcal{O}\left(\frac{\tilde z_i}{z_0}\right), \qquad S_\ell = \ell J_{\text{MG}} z_0^{\ell-1} + \mathcal{O}\left(\frac{\tilde z_i}{z_0}\right).\ee

\subsection{Large dipole geometries imply matching ratios}\label{sec:forward}
From the expressions (\ref{eq:multlargedipoleBH}) and (\ref{eq:multlargedipoleMG}) we derived above for  large dipole black holes and large dipole microstate geometries, it is clear that when the black hole and microstate geometry both have large dipoles, all finite multipole ratios \emph{of mass multipoles $M_\ell$} calculated in the \indirect or \direct ways will agree. Moreover, $M_\text{BH}/D$ and $\tilde z_i/z_0$ will give a quantitative estimate of the amount of mismatch between such mass multipole ratios. For example, consider the ratio:
\be \mathcal{M} \equiv \frac{M_2M_2}{M_4M_0}.\ee
Calculating this ratio using the \indirect method gives:
\be\label{eq:ratio2240indirectlargeD} \mathcal{M}_{\text{indirect}} = -\frac13 + \mathcal{O}\left(\left(\frac{D}{M}\right)^{-1}\right), \ee
whereas calculated using the \direct method gives:
\be \mathcal{M}_{\text{indirect}} = -\frac13 + \mathcal{O}\left(\frac{\tilde z_i}{z_0}\right).\ee

An analogous reasoning immediately tell us that multipole ratios involving only \emph{current} multipoles will also match when both black hole and microstate geometry have large dipoles. Less trivial is the matching of multipole ratios which involve both mass \emph{and} current multipoles. For example, consider $\mathcal{M}\equiv S_2/M_2$. For the large dipole black hole, we have:
\be \label{eq:lDipIndirectMom1} \mathcal{M} = 2 \frac{J_{\text{BH}}}{a D} + \mathcal{O}\left(\left(\frac{D}{M}\right)^{-1}\right),\ee
whereas for the large dipole microstate geometry we have:
\be \label{eq:lDipIndirectMom2} \mathcal{M} = -2 \frac{J_{\text{MG}}}{M_\text{BH}\,z_0} + \mathcal{O}\left(\frac{\tilde z_i}{z_0}\right),\ee
where we have replaced $M_{MG}=M_{BH}$, since the supersymmetric limit of the black hole in the \indirect method must be the same as the scaling black hole limit of the microstate geometry in the \direct method.
Thus, their ratios will match if and only if:
\be \label{eq:Jsmatching} \frac{J_\text{BH}}{aD} = -\frac{J_{\text{MG}}}{M_\text{BH}z_0} .\ee
Note that this is precisely the correspondence that we used for a two-center geometry\footnote{The first equality in \eqref{eq:JoveraD} is the identification of the black hole as a two-centre geometry, but it is also the expression for $-J_{MG}/(M_{MG}\,z_0)$ in a large dipole microstate.} in Section \ref{sec:BHtwocenter}! However, for a generic (multicenter) microstate geometry, (\ref{eq:Jsmatching}) seems to be an additional demand on the geometry.

The matching of other mixed mass and current multipole ratios is entirely analogous.
We conclude that multipole ratios between \indirect and \direct methods agree well if they are calculated for ``large dipole'' microstate geometries and ``large dipole'' black holes that additionally satisfy (\ref{eq:Jsmatching}); so we have shown that:
\be \begin{aligned} \label{eq:conjForward} \left(\frac{|D|}{M_\text{BH}}\gg 1\right) \text{ and } \left(\frac{z_0}{|\tilde z_i|}\gg 1\right) \text{ and } \left(\frac{J_\text{BH}}{aD} = -\frac{J_{\text{MG}}}{M_\text{BH}z_0}\right)\\
 \quad \Rightarrow \text{direct/indirect multipole ratios match}.
 \end{aligned}\ee

\subsection{Matching ratios imply large dipole geometries}\label{sec:3centres}

In this section, we will give evidence for the converse direction of \eqref{eq:conjForward}. We will consider a three-center, scaling geometry with center positions given by:
\begin{equation}
x_{1,2,3}=\mathcal{O}(\tilde\varepsilon),\quad z_1=z_0-\frac{\varepsilon}{2},\quad z_3=z_0+\frac{\varepsilon}{2},\quad\frac{z_{0,2}}{\varepsilon}\gg 1,\label{eq:2centreApprox}
\end{equation}
with small parameters $\varepsilon,\tilde\varepsilon>0$.
This assumption makes the geometry approximately axisymmetric, so that $x_i$ are small compared to the other length scales in the problem; as a result, working to zeroth order in $\tilde\varepsilon$, we are able to use the axisymmetric formulae \eqref{eq:MGmultipoles} for the multipoles.
Furthermore, we assume that centers $1$ and $3$ lie very close to each other, $z_3-z_1=\varepsilon\ll z_{1,3}$.

Demanding mass-centered coordinates, $M_1=0$, we can use \eqref{eq:MGmultipoles} to express
\begin{equation}
z_2=-\frac{\mm_1+\mm_3}{\mm_2}z_0+\varepsilon\frac{\mm_1-\mm_3}{2\,\mm_2} = -u\,z_0 + \mathcal{O}(\varepsilon).\label{eq:massCenteredConstr}
\end{equation}
where we have defined
\begin{equation}
u=\frac{\mm_1+\mm_3}{\mm_2},\label{eq:uSub}
\end{equation}
We are also assuming that
\begin{equation}\label{eq:assmp1}
\frac{\mm_1+\mm_3}{\mm_2}\sim\mathcal{O}(\epsilon^0),\quad \frac{\mm_1-\mm_3}{\mm_2}\sim\mathcal{O}(\epsilon^0).
\end{equation}
From (\ref{eq:massCenteredConstr}), this microstate geometry has a \emph{large dipole} as defined in Section \ref{sec:largeDipoleMS} when $u+1\ll 1$.

Further, from $S_0=0$ (which is equivalent to one of the bubble equations), we also have $\sm_1+\sm_2+\sm_3=0$. We can then express all of the multipoles \eqref{eq:MGmultipoles} of this geometry as
\begin{equation}
M_\ell=\mm_2\,z_0^\ell\,\Big(1-(-u)^{\ell-1}\Big)+\mathcal{O}(\frac{\varepsilon}{z_0}),\quad S_\ell=\sm_2\,z_0^\ell\,\Big(1+(-u)^\ell\Big)+ \mathcal{O}(\frac{\varepsilon}{z_0}).
\end{equation}

\paragraph{\Direct mass multipole ratios}
First, we consider a particular family of multiple ratios of mass multipoles from \cite{Bena:2020uup}; for our microstate geometry introduced above:
\begin{equation}
R_{1,MG}=\frac{M_{\ell+1}\,M_{\ell+2}}{M_\ell\,M_{\ell+3}}=\frac{\big(1-(-u)^\ell\big)\big(1-(-u)^{\ell+1}\big)}{\big(1-(-u)^{\ell-1}\big)\big(1-(-u)^{\ell+2}\big)}.\label{eq:massesOnlyMS}
\end{equation}

We start by investigating the structure of this ratio \eqref{eq:massesOnlyMS} as a function of $u$; its derivative is:
\begin{equation}
\frac{d\,R_{1,MG}}{du}=(-1)^\ell\,u^{\ell-2}(1-u^2)\frac{1-(1-u)^2(-u)^\ell-\ell(1+u)(1+u^{2\,\ell+1})+\Big(2u\frac{1-u^{2\,\ell-1}}{1-u}+u^{2\,\ell+1}\Big)}{(1-(-u)^{2+\ell})^2(1-(-u)^{\ell-1})^2}.
\end{equation}
This derivative vanishes at $u=\pm1$ and at $u=0$ for $\ell\geq 3$. Numerics suggest that the big fraction does not give any real roots. The values at the extrema are:
\begin{align}
\begin{array}{rcl}
R_{1,MG}=
\left\{\begin{array}{@{}l@{}}
1+\frac{2}{\ell^2+\ell-2},\quad\hfill u=-1\vspace{1mm}\\[\jot]
1,\quad\hfill u=0\vspace{1mm}\\[\jot]
1+\frac{3+(-1)^{\ell+1}(2\,\ell+1)}{\ell^2+\ell-2},\quad u=1\vspace{1mm}
\vspace{1mm}\end{array}\right.
\\
\end{array}\label{eq:massesOnlyMSAtExtrema}
\vspace*{7.5mm}.
\end{align}
The types of the extrema depend on the parity of $\ell$. Further, for large $|u|$, we have:
\begin{equation}
\lim\limits_{u\rightarrow\pm\infty}R_{1,MG}=1.
\end{equation}
We see that $R_{1,MS}$ squeezes 1 as $\ell$ grows larger.

\paragraph{\Indirect mass multipole ratios}
We now turn to the same multipole ratios, but for the general eight-charge black hole of (\ref{eq:MSgenBH}); they can be written as:
\be \label{eq:massesOnlyBH}
R_{1,BH}=\frac{M_{\ell+1}\,M_{\ell+2}}{M_\ell\,M_{\ell+3}}=\frac{(i+t)^{2\,\ell+1}+\big((-i+t)^{2\,\ell+1}-2t(1+t^2)^\ell\big)}{-2t(-3+t^2)(1+t^2)^{\ell-1}+\big((-i+t)^{2\,\ell+1}+(i+t)^{2\,\ell+1}\big)},\ee
where we have defined the shorthand
\be t=\frac{D}{M_{BH}}. \ee

We again analyze this ratio \eqref{eq:massesOnlyBH} as a function of $t$.
Its derivative is
\begin{align}
\label{eq:bhDervMassesOnly}\frac{d\,R_{1,BH}}{dt}&=\frac{8\,(1+t^2)^{\ell-2}}{\big[2\,t\,(3-t^2)(1+t^2)^{\ell-1}+H_+(t,2\,\ell+1)\big]^2}\\
&\times \big[i\,H_-(t,2\,\ell)+2\,t^3\big(H_+(t,2\,\ell)-2(1+t^2)^\ell\big)+2\,H_+(t,2\,\ell)\,t\,\ell+i\,t^2H_-(t,2\,\ell)(2\,\ell-1)\big],\notag\\
\nonumber H_{\pm}(t,n)&=\big((-i+t)^n\pm(i+t)^n\big),
\end{align}
The only obvious zeroes of the expression above are $\pm\,i$ which come from the factors of $(1+t^2)$. However, $t=0$ is also always a root. To see this, note that $H_{\pm}(t,2\,\ell)$ are of the form
\begin{align}
\begin{array}{rcl}
H_{\pm}(t,2\,\ell)=\rho^{2\,\ell}\pm\overline{\rho}^{2\,\ell}=r^{2\,\ell}(e^{i\,2\,\ell\,\theta}\pm e^{-i\,2\,\ell\,\theta})=
\left\{\begin{array}{@{}l@{}}
2\,r^{2\,\ell}\,\cos(2\,\ell\,\theta),\vspace{1mm}\\[\jot]
2\,i\,r^{2\,\ell}\,\sin(2\,\ell\,\theta),\vspace{1mm}
\vspace{-1mm}\end{array}\right.
\end{array}\quad\rho\in\mathbb{C}\label{eq:deMoivreInAction}.
\end{align}
The multi-argument trigonometric functions appearing in \eqref{eq:deMoivreInAction} can be converted to powers of single-angle sines and cosines with the help of the formulae in Appendix \ref{app:multAngles}. It is important to note that the expression for $\sin(n\,\theta)$ has an overall factor of $\sin\,\theta$, whereas $\cos(n\,\theta)$ does not. Combining these formulae with\footnote{We keep the $2$ in $2\,\ell\,\theta$ with the angle $\theta$ and identify $n$ in the Appendix \ref{app:multAngles} formulae with $\ell$.}
\begin{equation}
\sin(2\arctan(x))=\frac{2\,x}{1+x^2},\quad \cos(2\arctan(x))=\frac{1-x^2}{1+x^2},
\end{equation}
and identifying $\theta=\arctan(-1/t)$ (since $\rho=-i+t$) in \eqref{eq:deMoivreInAction}, we deduce that $H_-(t,2\,\ell)$ has a factor of $2\,t\,\ell$, whereas $H_+(t,2\,\ell)$ does not. The numerator of \eqref{eq:bhDervMassesOnly} then behaves as $\sim t^3$ when $t\rightarrow0$, because the first non-trivial terms between $i\,H_-(t,2\,\ell)$ and $2\,H_+(t,2\,\ell)\,t\,\ell$ cancel. Moreover, the square in the denominator goes as $t^2$ with $t\rightarrow0$. Thus, \eqref{eq:bhDervMassesOnly} has a single zero at $t=0$. Numerically, we additionally see that as $\ell$ grows, additional pairs of roots appear, symmetrically around $t=0$ and further and further away from the origin.\footnote{Although we do not discuss these roots further, we explicitly checked that the behaviour of $R_{1,BH}$ around these additional zeroes is such that it will never come close to matching with $R_{1,MG}(u)$ (for any value of $u$).}
Evaluating \eqref{eq:massesOnlyBH} at $t=0$,
\begin{equation}
\lim\limits_{t\rightarrow 0}R_{1,BH}=1-\frac{4}{3+(-1)^\ell(2\,\ell+1)},
\end{equation}
which agrees with \eqref{eq:massesOnlyMSAtExtrema} for $u=1$. The asymptotic behaviour of $R_{1,BH}$ is
\begin{equation}
\lim\limits_{t\rightarrow\pm\infty}R_{1,BH}=1+\frac{2}{\ell^2+\ell-2},\label{eq:R1BHtinf}
\end{equation}
which agrees with \eqref{eq:massesOnlyMSAtExtrema} for $u=-1$.

\paragraph{Beyond only mass multipole ratios}
The analysis above suggests that there are only two points where the mass multipole ratios $R_{1,BH}$ and $R_{1,MG}$ \emph{can} agree for arbitrary $\ell$:
\begin{itemize}
 \item $t\rightarrow0$ and $u\rightarrow1$, or
 \item $t\rightarrow\pm\infty$ with $u\rightarrow-1$.
\end{itemize}
(We have checked explicitly for values of $\ell$ up to $50$ that these are indeed the only points of agreement for these mass multipole ratios.)

To distinguish which of these cases will give us matching multipole ratios for \emph{all} types of ratios, we can consider a mixed multipole ratio such as (again from \cite{Bena:2020uup}):
\begin{align}
R_{2,MG}&=\frac{S_{\ell+1}\,S_{\ell+2}}{M_\ell\,M_{\ell+3}}=\Big(\frac{\sm_2}{\mm_1+\mm_3}\Big)^2\frac{\big(1-(-u)^{1+\ell}\big)}{\,\big(1-(-u)^{-1+\ell}\big)},\notag\\
\label{eq:mixedratios} R_{2,BH}&=\frac{S_{\ell+1}\,S_{\ell+2}}{M_\ell\,M_{\ell+3}}=\Big(\frac{J_{BH}}{a\,D}\Big)^2\frac{\bigg[\Big(-\frac{i}{t}+1\Big)^{1+\ell}-\Big(\frac{i}{t}+1\Big)^{1+\ell}\bigg]}{\Big(\frac{1}{t^2}+1\Big)^2\bigg[\Big(-\frac{i}{t}+1\Big)^{\ell-1}-\Big(\frac{i}{t}+1\Big)^{\ell-1}\bigg]},
\end{align}
First, consider the scenario where $t\rightarrow0$ and $u\rightarrow1$. Take $u=1+\mathcal{O}(\delta)$ and $t=\mathcal{O}(\tilde{\delta})$, with $\delta,\tilde\delta$ small (and a priori not the same) parameters. Expanding to leading order in these parameters gives:
\begin{align}
R_{2,MG}&=\frac{(-1)^\ell\big((-1)^\ell\,\ell-1\big)}{\ell-1}\Big(\frac{\sm_2}{\mm_1+\mm_3}\Big)^2+\mathcal{O}(\delta),\notag\\
R_{2,BH}&=-\frac{(-1)^\ell\big((-1)^\ell\,\ell-1\big)}{\ell-1}\Big(\frac{\nu_1}{\mu_2}\Big)^2+\mathcal{O}(\tilde{\delta}).
\end{align}
Clearly, in this limit we are \emph{not} able to recover matching multipole ratios.

For the second scenario, where $t\rightarrow\pm\infty$ and $u\rightarrow-1$, take $u=-1+\delta$ and $t=\tilde{\delta}^{-1}$ and again expand to leading order:\footnote{Note that we assume $\sm_2/(\mm_1+\mm_3)\gg\delta$ and $J_\text{BH}/(aD) \gg \tilde\delta$ here.}
\begin{align}
R_{2,MG}&=\frac{\ell+1}{\ell-1}\Big(\frac{\sm_2}{\mm_1+\mm_3}\Big)^2+\mathcal{O}(\delta),\notag\\
R_{2,BH}&=\frac{\ell+1}{\ell-1}\Big(\frac{J_{BH}}{a\,D}\Big)^2+\mathcal{O}(\tilde{\delta}).\label{eq:match2ndScen}
 \end{align}
With that, for the two expressions in \eqref{eq:match2ndScen} to match, one clearly needs
\begin{equation}
\frac{\sm_2}{\mm_1+\mm_3}=\pm\frac{J_{BH}}{a\,D}.\label{eq:angMomMatch}
\end{equation}

Using \eqref{eq:massCenteredConstr}, \eqref{eq:uSub}, $\sm_1+\sm_2+\sm_3=0$, and $M_{MG}=\mm_1+\mm_2+\mm_3$, we can express the microstate geometry angular momentum as:
\begin{equation}
J_{MG}=\sm_1\,z_1+\sm_2\,z_2+\sm_3\,z_3=\frac{\sm_2\,z_0\,M_{MG}}{\mm_1+\mm_3}+\mathcal{O}(\epsilon,\delta),
\end{equation}
We can then rewrite (\ref{eq:angMomMatch}) as:
\begin{equation}\label{eq:JJrewrite}
\frac{J_{MG}}{z_0\,M_{MG}}=\frac{\sm_2}{\mm_1+\mm_3}=\pm\frac{J_{BH}}{a\,D},
\end{equation}
up to terms subleading in $\epsilon,\delta$. The sign can further be determined by considering a ratio such as $S_2/M_2$, which fixes the sign to be the \textit{minus}.

Any other types of multipole ratios will not provide any additional information; they all agree if the conditions outlined above are satisfied.

\paragraph{Summary: Necessary matching conditions}
From the above considerations, if the microstate geometry mass multipole ratios (\ref{eq:massesOnlyMS}) and the black hole mass multipole ratios (\ref{eq:massesOnlyBH})  are to match, \emph{as well as} the mixed multipole ratios (\ref{eq:mixedratios}), it is clear that we require the limit $u\rightarrow -1$ for the microstate geometries and $t\rightarrow\pm\infty$ for the black hole ratios --- these conditions are precisely the large dipole conditions\footnote{$z_2=z_0+\mathcal{O}(\varepsilon,\delta)$ from \eqref{eq:massCenteredConstr} with $u\rightarrow-1$.} from Section \ref{sec:largeDipoleMS}, resp. $z_0/\tilde z_i\gg 1$ and $|D|/M\gg 1$.

Moreover, we have found that the mixed multipole ratios matching indeed implies that (\ref{eq:Jsmatching}) must hold. 



In summary, for the particular kind of three-center, scaling microstate geometry that we have discussed here, we can conclude that:
\be \begin{aligned} \label{eq:conjBackward}
\text{direct/indirect multipole ratios match} \Rightarrow\\
\left(\frac{|D|}{M_\text{BH}}\gg 1\right) \text{ and } \left(\frac{z_0}{|\tilde z_i|}\gg 1\right) \text{ and } \left(\frac{J_\text{BH}}{aD} = -\frac{J_{\text{MG}}}{M_\text{BH}z_0}\right),
 \end{aligned}\ee
 which is the converse of (\ref{eq:conjForward}).

\section{Examples}\label{sec:examples}

In this section, we give a number of examples that illustrate our arguments above on concrete microstate geometries. In Section \ref{sec:limitMG}, we give an explicit realization of a three-center geometry such as used in Section \ref{sec:3centres}. In Section \ref{sec:excounter}, we give an explicit example that shows that $\mathcal{H}\ll 1$ is not sufficient to guarantee matching of multipole ratios (giving a counterexample to the original conjecture of \cite{Bena:2020see,Bena:2020uup}). Finally, in Section \ref{sec:geometriesAB}, we give the geometries $A,B$ from \cite{Bena:2020uup}, which have matching multipole ratios, and explain how they fit within our framework.

\subsection{Limit geometry}\label{sec:limitMG}

In this section we give an explicit example of a two-parameter family of three-centre geometries that in the scaling limit have a regime in which their multipole ratios match with the corresponding BH. Without loss of generality, we take the three centres to be in the $x-z$ plane and demand the centre of mass to be at the origin (i.e. $M_1=0$). We also align the angular momentum with the $z$-axis.

We work with the following moduli:
\begin{equation}
\big(v_0,\,k^1_0,\,k^2_0,\,k^3_0,\,l^1_0,\,l^2_0,\,l^3_0,\,m_0\big)=\big(1,0,0,0,1,1,1,0\big).
\end{equation}
The $v_i$ and $k^{\hat{I}}_i$ center charges are given in terms of two tuneable parameters $\nu,\varphi$:
\begin{align}
v_i&=\big(1,\nu,-\nu\big),\notag\\
k^1_i&=\Bigg(\frac{\varphi\,(1-9\,\varphi^2)}{3(1+\varphi^2)},\varphi,\varphi\Bigg),\\
k^2_i&=\Bigg(-\frac{2\,\nu\,\varphi\big(\varphi^2(2-3\,\varphi^2)+\nu(1-2\,\varphi^2-3\,\varphi^4)+6\,\nu^2(\varphi^4-1)\big)}{3(1+\varphi^2)(4\,\nu^2(1+\varphi^2)-\varphi^2)},\frac{\varphi}{2}\big(2\,\nu-1\big),\frac{\varphi}{2}\big(2\,\nu-1\big)\Bigg),\notag\\
k^3_i&=\Bigg(\frac{2\,\nu\,\varphi\big(\varphi^2(2-3\,\varphi^2)+\nu(3\,\varphi^4+2\,\varphi^2-1)+6\,\nu^2(\varphi^4-1)\big)}{3(1+\varphi^2)(4\,\nu^2(1+\varphi^2)-\varphi^2)},-\frac{\varphi}{2}\big(2\,\nu+1\big),-\frac{\varphi}{2}\big(2\,\nu+1\big)\Bigg).\notag
\end{align}
where $\nu>(\sqrt{3}/4)\,\varphi$; all three centers are smooth, so the other center charges are determined by (\ref{eq:MGsmoothness}). The electric and magnetic charges of the geometry are given by:
\begin{align}\label{eq:MGcharges}
\big(Q_0,\,Q_1,\,Q_2,\,Q_3\big)&=\big(1,\varphi^2,\varphi^2,\varphi^2\big),\\
\big(P^0,\,P^1,\,P^2,\,P^3\big)&=\Bigg(\frac{\varphi^3(1-9\,\varphi^2+24\,\nu^2(1+\varphi^2))}{3(1+\varphi^2)(4\,\nu^2(1+\varphi^2)-\varphi^2)},-\frac{\varphi^3(1-9\,\varphi^2+24\,\nu^2(1+\varphi^2))}{3(1+\varphi^2)(4\,\nu^2(1+\varphi^2)-\varphi^2)},0,0\Bigg),\notag
\end{align}
so that the entropy parameter is given by:
\begin{equation}
\mathcal{H}=\frac{(3\,\varphi^2-1)(1-15\,\varphi^2+48\,\nu^2(1+\varphi^2))}{36(\varphi^2-4\,\nu^2(1+\varphi^2))^2}.
\end{equation}
The bubble equations can be seen as determining the intercenter distances in terms of a single tuneable parameter which we take to be $r_{23}=L$. The scaling limit is then $L\rightarrow0$.

In the scaling limit, the matching of multipole ratios happens for $\nu\rightarrow\infty$; note that in this limit, the entropy parameter and magnetic charges are:
\begin{equation}\label{eq:entropyParLimGeo}
\lim\limits_{\nu\rightarrow\infty}\mathcal{H}=\frac{3\,\varphi^2-1}{12\,\nu^2(1+\varphi^2)}+\mathcal{O}(\nu^{-4}),\quad \lim\limits_{\nu\rightarrow\infty}P^0=-\lim\limits_{\nu\rightarrow\infty}P^1=\frac{2\,\varphi^3}{1+\varphi^2}+\mathcal{O}(\nu^{-2}).
\end{equation}
Also in the scaling limit at large $\nu$, the center positions are given by:
\begin{align}\label{eq:limGeoPos}
x_1&=L\,\Bigg[\frac{\varphi\,(7-4\,\varphi^2+\varphi^4)}{\sqrt{3}\,(3+\varphi^4)(3\,\varphi^2-1)}+\mathcal{O}(\nu^{-2})\Bigg]+\mathcal{O}(L^2),\notag\\
x_2&=L\,\Bigg[-\frac{2\,\varphi\,(1+\varphi^2)^2}{\sqrt{3}\,(3+\varphi^4)(3\,\varphi^2-1)}-\frac{\varphi\,(7-4\,\varphi^2+\varphi^4)}{2\,\sqrt{3}\,\nu\,(3+\varphi^4)(3\,\varphi^2-1)}+\mathcal{O}(\nu^{-2})\Bigg]+\mathcal{O}(L^2),\notag\\
x_3&=L\,\Bigg[-\frac{2\,\varphi\,(1+\varphi^2)^2}{\sqrt{3}\,(3+\varphi^4)(3\,\varphi^2-1)}+\frac{\varphi\,(7-4\,\varphi^2+\varphi^4)}{2\,\sqrt{3}\,\nu\,(3+\varphi^4)(3\,\varphi^2-1)}+\mathcal{O}(\nu^{-2})\Bigg]+\mathcal{O}(L^2),\notag\\
z_1&=L\,\Bigg[\frac{\nu\,(1-\varphi^2)^2}{(1+\varphi^2)(1+3\,\varphi^2)}-\frac{f_1(\varphi)}{36\,\nu\,(1-3\,\varphi^2)(1+\varphi^2)^2(3+\varphi^4)^2}+\mathcal{O}(\nu^{-2})\Bigg]+\mathcal{O}(L^2),\notag\\
z_2&=L\,\Bigg[\frac{\nu\,(1-\varphi^2)^2}{(1+\varphi^2)(1+3\,\varphi^2)}-\frac{1}{2}-\frac{f_2(\varphi)}{72\,\nu\,(1+\varphi^2)^2(3+\varphi^4)^2}+\mathcal{O}(\nu^{-2})\Bigg]+\mathcal{O}(L^2),\notag\\
z_3&=L\,\Bigg[\frac{\nu\,(1-\varphi^2)^2}{(1+\varphi^2)(1+3\,\varphi^2)}+\frac{1}{2}-\frac{f_2(\varphi)}{72\,\nu\,(1+\varphi^2)^2(3+\varphi^4)^2}+\mathcal{O}(\nu^{-2})\Bigg]+\mathcal{O}(L^2),
\end{align}
with
\begin{align}
f_1(\varphi)&=270\,\varphi^{16}+333\,\varphi^{14}+1371\,\varphi^{12}+429\,\varphi^{10}+2175\,\varphi^8+303\,\varphi^6+905\,\varphi^4+663\,\varphi^2+79,\notag\\
f_2(\varphi)&=69\,\varphi^{12}+66\,\varphi^{10}+291\,\varphi^8+108\,\varphi^6+235\,\varphi^4+114\,\varphi^2+77.
\end{align}
The parameter $u$ from Section \ref{sec:3centres} is given by
\begin{align}\label{eq:uLim}
u=\frac{\mm_1+\mm_2}{\mm_3}=-1-\frac{(1+\varphi^2)(1+3\,\varphi^2)}{\nu\,(1-\varphi^2)^2}+\mathcal{O}(\nu^{-2}).
\end{align}
The relabelling of indices comes about because centres $(1,2,3)$ in Section \ref{sec:3centres} correspond to $(2,3,1)$ in \eqref{eq:limGeoPos}. In addition, we identify $z_0-L/4$ with the first two terms of $z_2$ in \eqref{eq:limGeoPos} and $P^0=-P^1$.

From the above, it is clear that this geometry (for large $\nu$) is an explicit realization of the three-center, almost-axisymmetric scaling geometry discussed in Section \ref{sec:3centres}, where we can identify $\nu\sim\frac{1}{\varepsilon}$ in (\ref{eq:2centreApprox}).

Since the geometry is approximately axisymmetric, we expect the multipole moments to be approximately given by the axisymmetric formulae (\ref{eq:MGmultipoles}). Explicitly, we find that:\footnote{The multipoles for a general, non-axisymmetric microstate geometry can be found in \cite{Bena:2020uup, Bianchi:2020miz}.}
\begin{align}\label{eq:MlmSlmLOexact}
M_{\ell m}^{MG}&=\Big[\frac{L\,\nu\,(1-\varphi^2)^2}{(1+\varphi^2)(1+3\,\varphi^2)}\Big]^\ell\nu^{-m}\,C^{(1)}_{\ell m}(\varphi)+\mathcal{O}(\nu^{\ell-m-1})+\mathcal{O}(L^{\ell+1}),\notag\\
S_{\ell m}^{MG}&=\Big[\frac{L\,\nu\,(1-\varphi^2)^2}{(1+\varphi^2)(1+3\,\varphi^2)}\Big]^\ell\nu^{-m}\,C^{(2)}_{\ell m}(\varphi)+\mathcal{O}(\nu^{\ell-m-1})+\mathcal{O}(L^{\ell+1}),
\end{align}
where the $C^{(1,2)}_{\ell m}(\varphi)$ are functions given in Appendix \ref{app:CfuncsMG}. The multipoles for $-\ell\leq m<0$ can be obtained via the relation $M_{\ell(-m)}^{MG}=(-1)^mM_{\ell m}^{MG}$. Clearly, the non-axisymmetric $m\neq 0$ terms are subleading with respect to the $m=0$ multipoles. We will restrict ourselves to the ($m=0$) axisymmetric multipoles in the following.

Note that we have:
\begin{align}
M_\ell^{(MG)}&=-\frac{1}{4}(\ell-1)(1+3\,\varphi^2)\Bigg(\frac{L\,\nu\,(1-\varphi^2)^2}{(1+\varphi^2)(1+3\,\varphi^2)}\Bigg)^\ell+\mathcal{O}(\nu^{\ell-1})+\mathcal{O}(L^{\ell+1}),\notag\\
S_\ell^{(MG)}&=\frac{\ell}{4}\frac{\varphi\,(1+3\,\varphi^2)(3+\varphi^4)}{(1-\varphi^2)^2}\Bigg(\frac{L\,\nu\,(1-\varphi^2)^2}{(1+\varphi^2)(1+3\,\varphi^2)}\Bigg)^\ell+\mathcal{O}(\nu^{\ell-1})+\mathcal{O}(L^{\ell+1}).
\end{align}
so that the ratios can be computed as:\footnote{We have checked that these expressions (to leading order in $\nu$) stay the same if instead of $M_\ell,S_\ell$, we use the quadratic invariants (valid for generic non-axisymmetric spacetimes) given by:
\begin{equation}
\mathfrak{M}^{(MG)}_{\ell}=\sqrt{\sum_{m=-\ell}^{\ell}\lvert M^{(MG)}_{\ell m}\rvert^2},\quad \mathfrak{S}^{(MG)}_{\ell}=\sqrt{\sum_{m=-\ell}^{\ell}\lvert S^{(MG)}_{\ell m}\rvert^2}.
\end{equation}
}
\begin{align}\label{eq:m0Ratios}
\frac{M^{(MG)}_{\ell+2}\,S^{(MG)}_{\ell}}{M^{(MG)}_\ell\,S^{(MG)}_{\ell+2}}&=1+\frac{2}{\ell^2+\ell-2}+\mathcal{O}(\nu^{-2})+\mathcal{O}(L),\notag\\
\frac{S^{(MG)}_{\ell+1}\,S^{(MG)}_{\ell+2}}{M^{(MG)}_\ell\,M^{(MG)}_{\ell+3}}&=\frac{\ell+1}{\ell-1}\frac{\varphi^2\,(3+\varphi^4)^2}{(1-\varphi^2)^4}+\mathcal{O}(\nu^{-2})+\mathcal{O}(L),\notag\\
\frac{M^{(MG)}_{\ell+1}\,M^{(MG)}_{\ell+2}}{S^{(MG)}_\ell\,S^{(MG)}_{\ell+3}}&=\frac{\ell+1}{\ell+3}\frac{(1-\varphi^2)^4}{\varphi^2\,(3+\varphi^4)^2}+\mathcal{O}(\nu^{-2})+\mathcal{O}(L).
\end{align}

We can compare the above results for our \direct method of calculating multipole ratios to the multipole ratios calculated using the \indirect method as the supersymmetric limit of non-extremal black hole ratios (as outlined in Section \ref{sec:2ways}). In practice, we follow the same procedure as in \cite{Bena:2020uup,Bena:2020see}: we take a non-extremal, rotating STU BH with the same asymptotic charges \eqref{eq:MGcharges} as the microstate geometry, and numerically approach the extremal, non-rotating limit. In that limit the BH's mass gets locked to the charges and we have $M_{BH}=M_{MG}=M$.

\begin{figure}[!ht]
	\centering
	\includegraphics[width=0.5\textwidth]{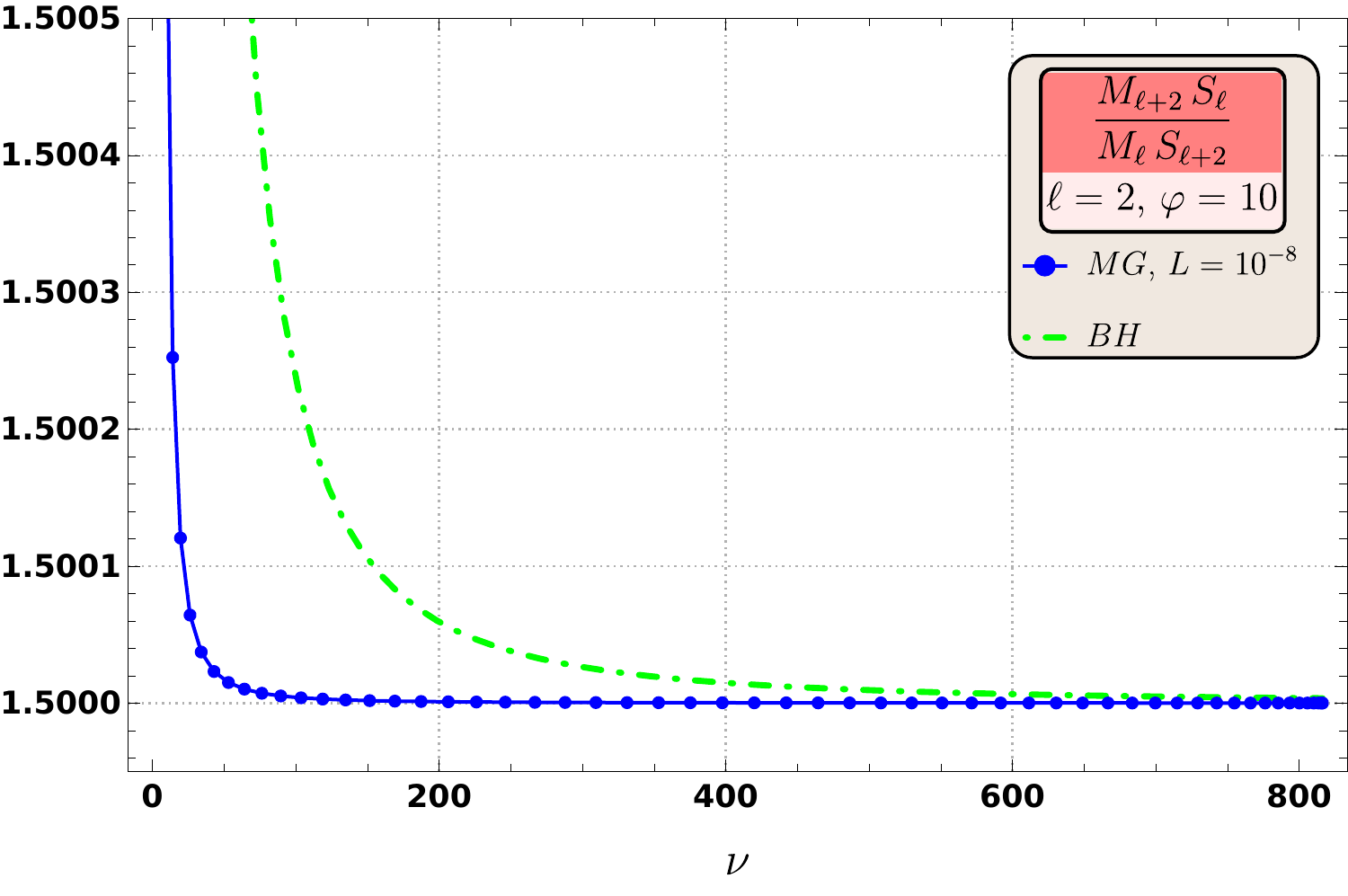}\hspace*{1mm}
	\includegraphics[width=0.51\textwidth]{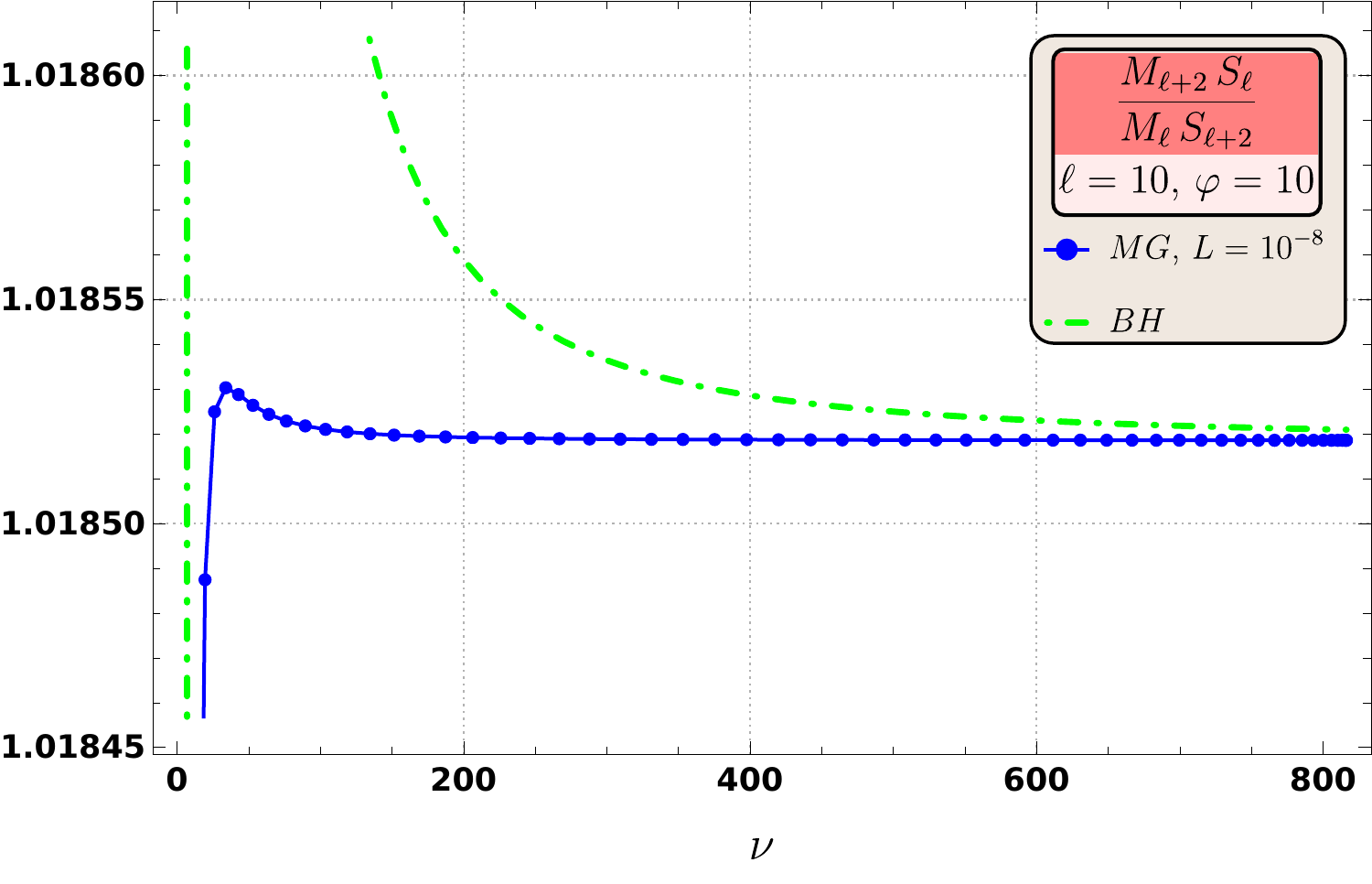}
	\caption{$(M_{\ell+2}\,S_{\ell})/(M_\ell\,S_{\ell+2})$ as a function of $\nu$ for BH and MG with same asymptotic charges \eqref{eq:MGcharges} with $\varphi=10$, $L=10^{-8}$, for $\ell=2,\,10$. $m=0$ case for MG multipoles.}
	\label{fig:MSMSm0comp}
\end{figure}

\begin{figure}[!ht]
	\centering
	\includegraphics[width=0.5\textwidth]{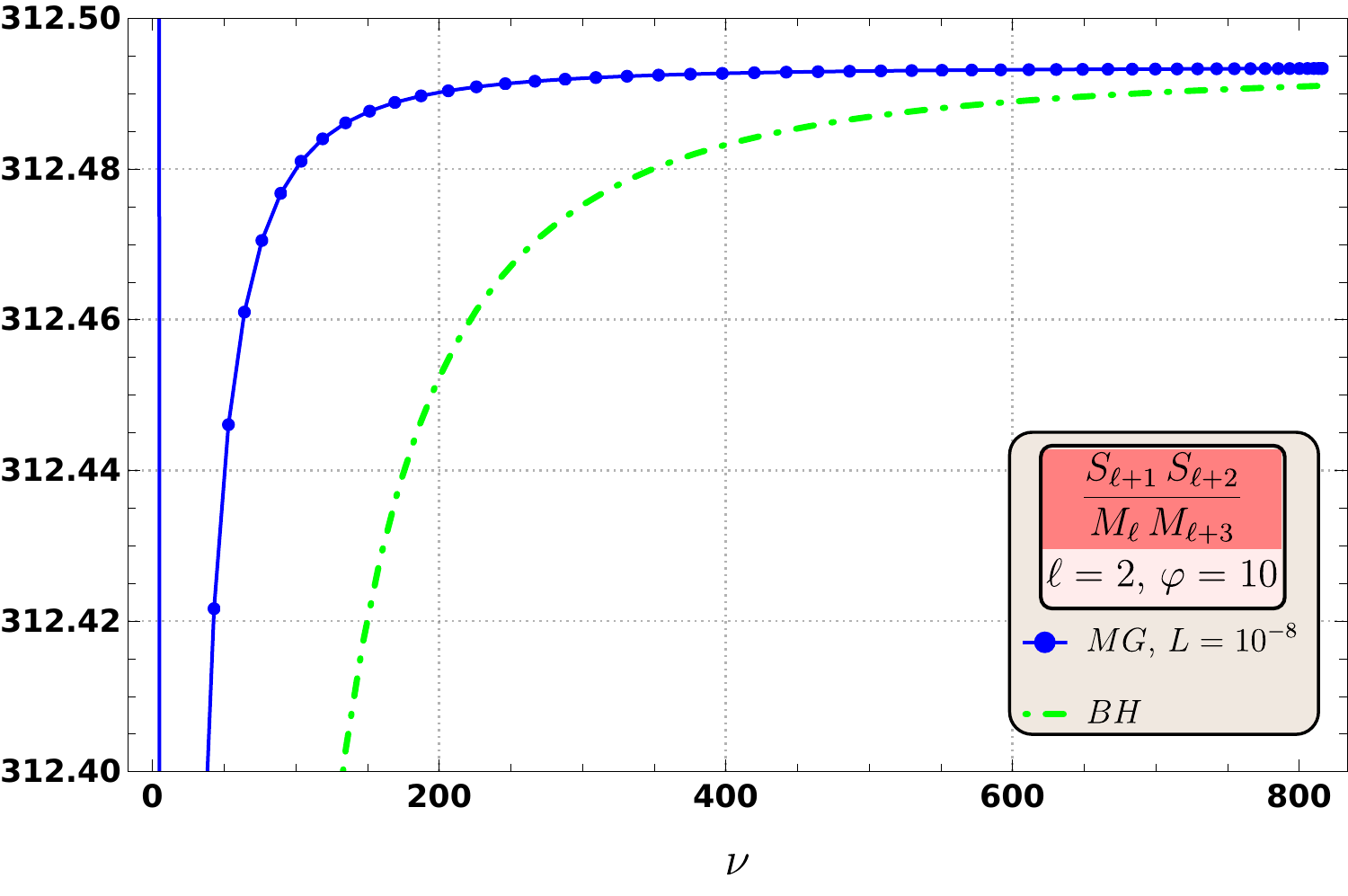}\hspace*{1mm}
	\includegraphics[width=0.51\textwidth]{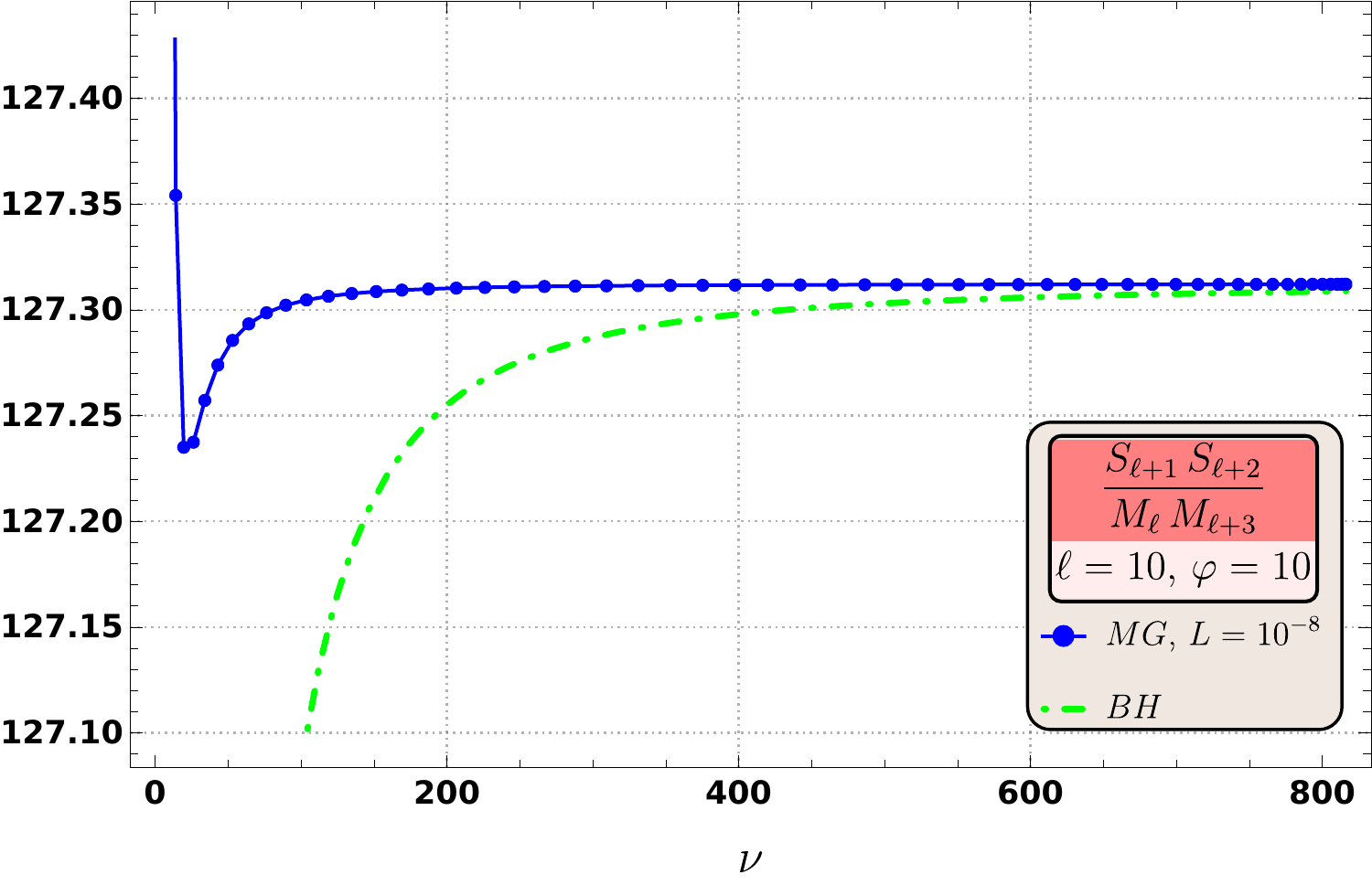}
	\caption{$(S_{\ell+1}\,S_{\ell+2})/(M_\ell\,M_{\ell+3})$ as a function of $\nu$ for BH and MG with same asymptotic charges \eqref{eq:MGcharges} with $\varphi=10$, $L=10^{-8}$, for $\ell=2,\,10$. $m=0$ case for MG multipoles.}
	\label{fig:SSMMm0comp}
\end{figure}

\begin{figure}[!ht]
	\centering
	\includegraphics[width=0.5\textwidth]{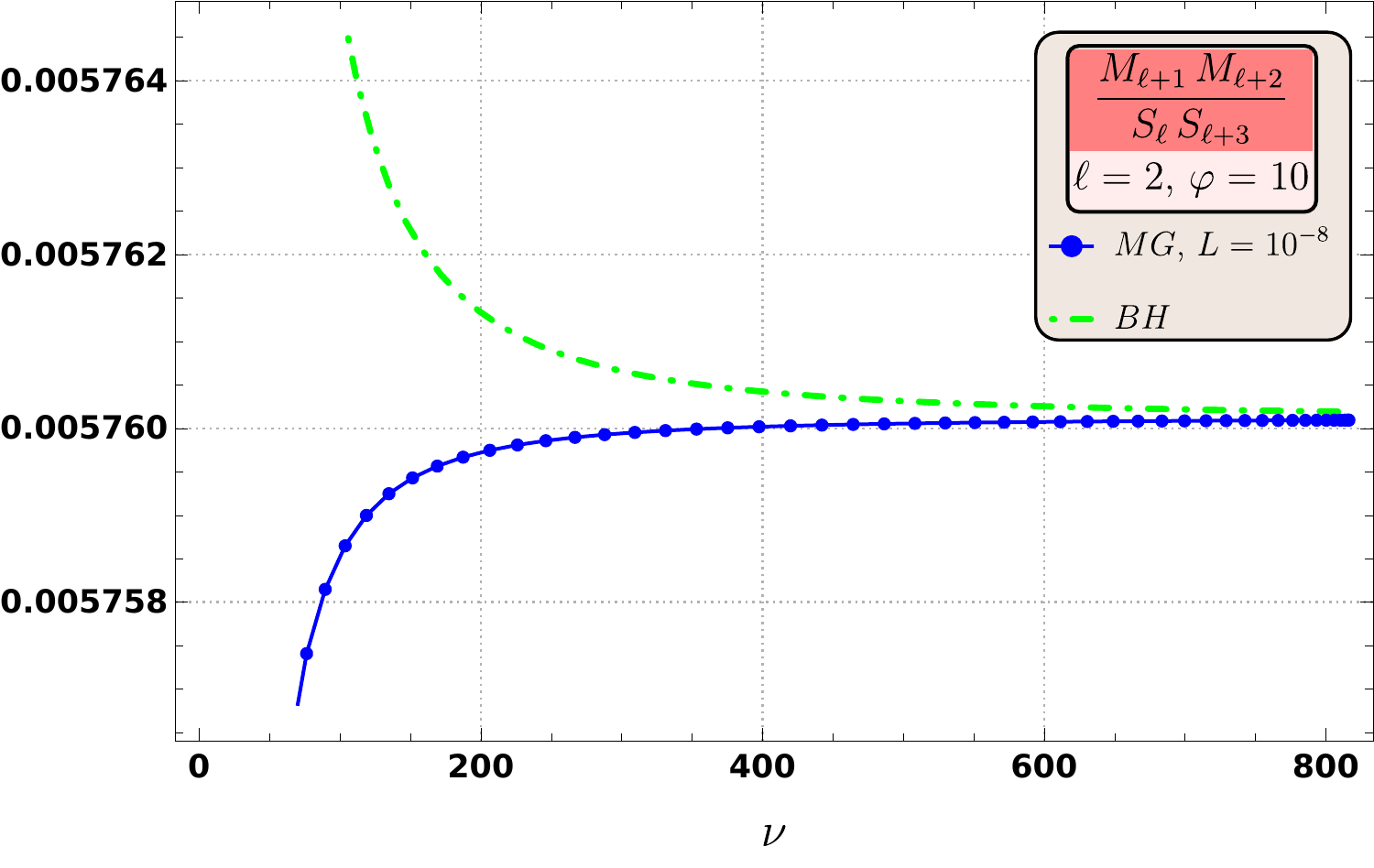}\hspace*{1mm}
	\includegraphics[width=0.51\textwidth]{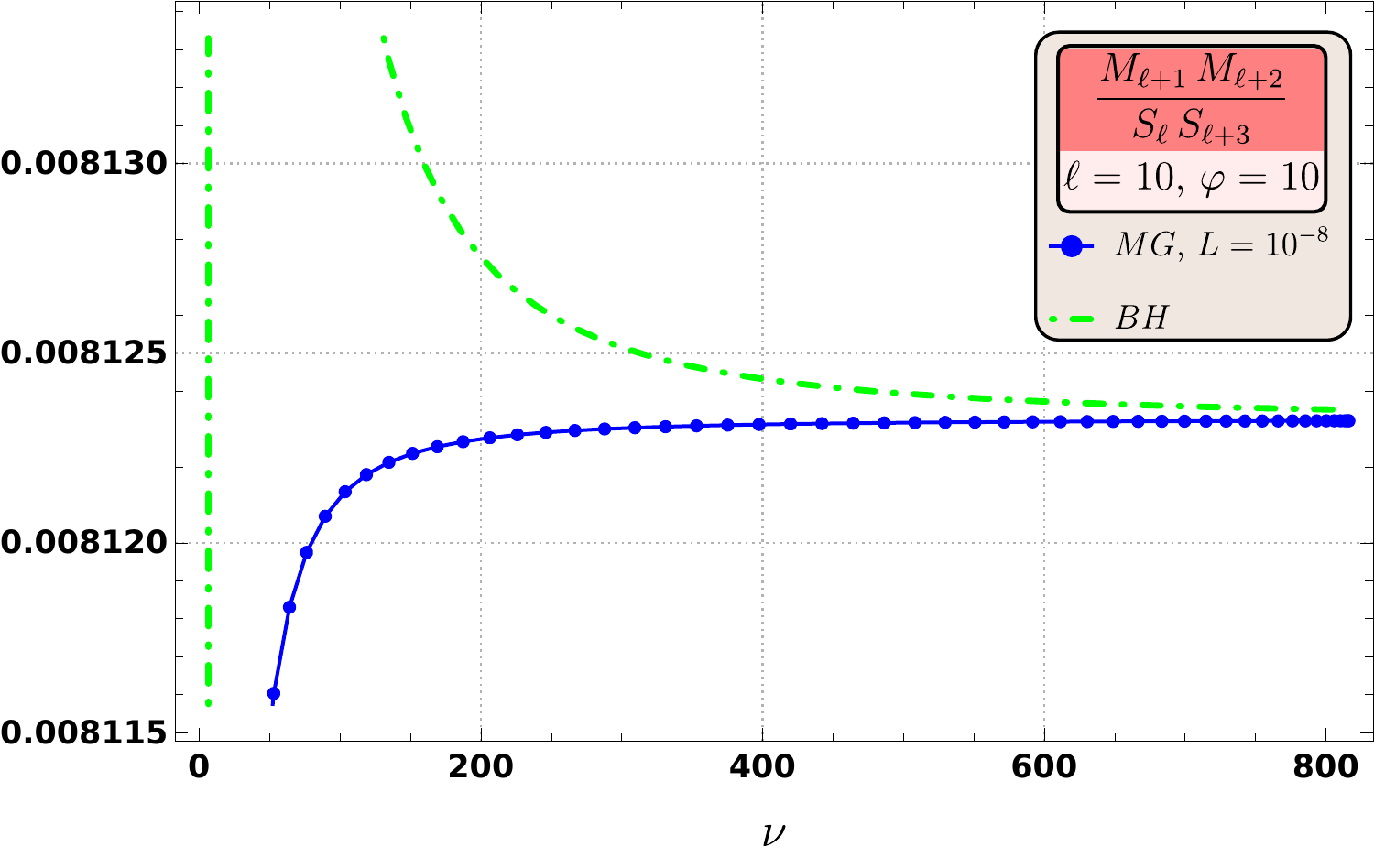}
	\caption{$(M_{\ell+1}\,M_{\ell+2})/(S_\ell\,S_{\ell+3})$ as a function of $\nu$ for BH and MG with same asymptotic charges \eqref{eq:MGcharges} with $\varphi=10$, $L=10^{-8}$, for $\ell=2,\,10$. $m=0$ case for MG multipoles.}
	\label{fig:MMSSm0comp}
\end{figure}

We show the ratios' comparison explicitly for $\varphi=10$ in Figures~\ref{fig:MSMSm0comp}, \ref{fig:SSMMm0comp} and \ref{fig:MMSSm0comp}, although the matching holds for any (allowed) value of $\varphi$.\footnote{Positivity of intercenter distances implies a minimum bound on $\varphi$. Values below it are excluded.} We show the multipole ratios as a function of $\nu$ (which dials the magnetic charges) to show explicitly how $\nu\rightarrow\infty$ is indeed necessary for matching of the multipole ratios. Note that we take $L=10^{-8}$ for the microstate geometry, which is sufficiently ``close'' to the scaling limit. 


As a consistency check we can look at \eqref{eq:Jsmatching}, which we claim is required to hold for the matching of ratios. Using \eqref{eq:JJrewrite}, accounting for the identification of centres as explained after \eqref{eq:uLim}, we obtain
\begin{equation}\label{eq:Jz0M}
-\frac{J_{MG}}{z_0\,M_{MG}}=-\frac{\sm_3}{\mm_1+\mm_2}=\frac{\varphi\,(3+\varphi^4)}{(1-\varphi^2)^2}+\mathcal{O}(\nu^{-1}),
\end{equation}
The above expression is to be compared with the last part of \eqref{eq:JoveraD}, where we take the electric charges from \eqref{eq:MGcharges}, as we set them equal for both geometries in the matching procedure. Importantly, \eqref{eq:JoveraD} is in the regime $Q_0\ll Q_{\hat{I}}$, which translates to $\varphi\gg1$. Applying that limit to \eqref{eq:Jz0M} gives exactly the same result as the RHS of \eqref{eq:MGcharges} - namely just $\varphi$. We have also confirmed that check using numerical results.

\begin{figure}[!ht]
	\centering
	\includegraphics[width=0.5\textwidth]{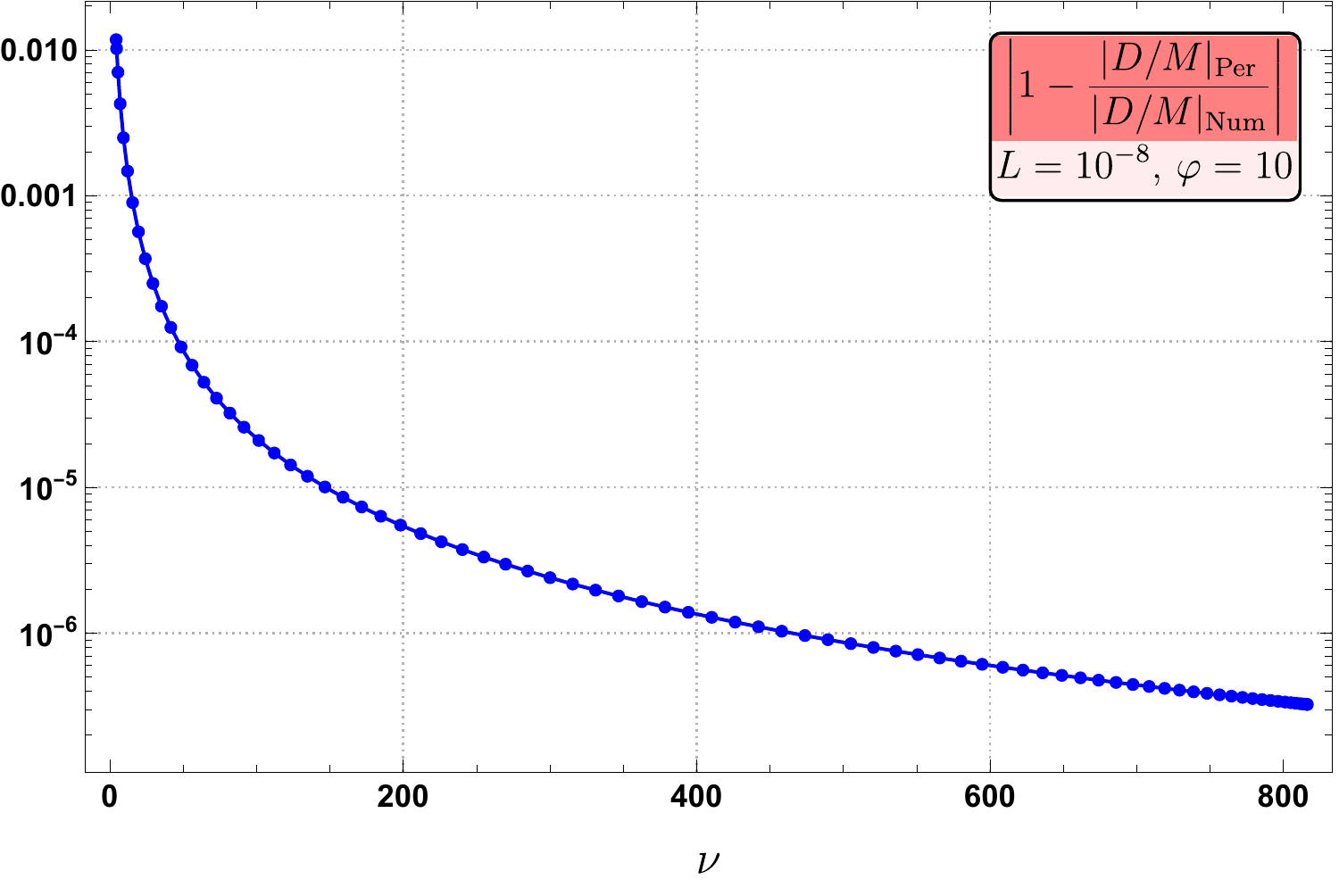}
	\caption{Log plot of the accuracy of the approximation for $D/M_{BH}$, \eqref{eq:DoverM} for \eqref{eq:MGcharges}, given by \eqref{eq:DMapproxLimitGeo}, against the exact numerical result with $\varphi=10$, $L=10^{-8}$.}
	\label{fig:DMp}
\end{figure}

Repeating the same operation for \eqref{eq:DoverM} gives us
\begin{equation}\label{eq:DMapproxLimitGeo}
\left\lvert\frac{D}{M_{BH}}\right\rvert=\left\lvert\frac{2\,\sqrt{3\,(1+\varphi^2)}\,(1-\varphi^2)^2}{\sqrt{3\,\varphi^2-1}\,(1+\varphi^2)(1+3\,\varphi^2)}\nu\right\rvert+\mathcal{O}(\nu^{-1}),
\end{equation}
where we replaced $\epsilon$ using its definition in \eqref{eq:P0parameps} and used the electric and magnetic charges in \eqref{eq:MGcharges}. This expression agrees excellently with the numerics as exhibited by Figure~\ref{fig:DMp}.

\subsection{Small dipole counterexample}\label{sec:excounter}

Here we give an explicit geometry that shows a small entropy parameter $\mathcal{H}\ll 1$ for the black hole is not sufficient to ensure the matching of multipole ratios as calculated with the \direct and \indirect methods; this provides a counterexample to the conjecture of \cite{Bena:2020see,Bena:2020uup} and shows the necessity of the additional condition that also the microstate geometry has a large dipole (as defined and discussed in Section \ref{sec:matchingproofs}).

We consider a smooth three-center scaling geometry with center charges $(v_i, k^I_i)$:
\be
\begin{aligned}
v_1 &= 1, & v_2& = -1, & v_3 &= 1,\\
k^1_1& = 2000.00099999975, & k^1_2 &= -1000 & k^1_3 &= 999.996,\\
k^2_1& = -1000.00099999575, & k^2_2 &= 1500.000499998375,\\
k^3_1 &=k^2_1 & k^3_2 &= k^2_2,
\end{aligned}
\ee
and $k^{\hat{I}}_3$ for ${\hat{I}}=2,3$ can be determined by the conditions $\sum_{i} k^{\hat{I}}_i= 0$ for ${\hat{I}}=2,3$. The $l^{\hat{I}}_i, m_i$ charges are determined by the smoothness conditions (\ref{eq:MGsmoothness}). The charges of the solution are:
\be Q_0 = 1, \qquad Q_{1,2,3} = 10^6, \qquad P^0 = -1999.997,\ee
so that the entropy parameter is tuned to be very small:
\be \mathcal{H} = 10^{-6},\ee
so the corresponding black hole has a large dipole $D$, in particular:
\be \label{eq:counterex-largeD} \frac{D}{M} = 333.332.\ee
We also note the black hole quantity $J/(Ma)$ for reference:
\be \frac{J}{aD} = 1000.0025.\ee

Near the scaling point, the center positions are (to leading order in the scaling parameter $\lambda\ll 1$):
\be \begin{aligned}
  x_1 &= \lambda, & z_1 &= -0.6299\, \lambda,\\
  x_2 &= 0.5\, \lambda, & z_2 &= 377961.6520\, \lambda,\\
  x_3 &= -3.5\, \lambda, & z_3 &= 377966.187552\,\lambda.
 \end{aligned}\ee
 Note that these center positions have already been chosen such that $M_1=0$ and the angular momentum $S_1=J$ is aligned with the $z$-axis.
Clearly, this three-center geometry is to a good approximation axisymmetric (as the $x$-coordinates are much smaller compared to the overall scale of the geometry). However, this geometry also clearly \emph{does not} fit in the large dipole definition (\ref{eq:largedipoleMG}), essentially since center 1 is near the origin.

Since the corresponding supersymmetric black hole has a large dipole (\ref{eq:counterex-largeD}), the multipole ratios for the black hole calculated using the \indirect way will approximately give the large dipole values such as those calculated in Section \ref{sec:forward}, for example (\ref{eq:ratio2240indirectlargeD}):
\be \mathcal{M}_{ind} = \frac{M_2M_2}{M_4M_0} \approx -\frac13.\ee
However, when calculated using the \direct method in the scaling limit of this microstate geometry, we find:
\be \mathcal{M}_{dir} = \frac{M_2M_2}{M_4M_0} = 3.1555\times 10^{-6}.\ee
Other multipole ratios can be calculated similarly; in general, no multipole ratios will match between the \indirect and \direct methods for this microstate geometry, showing the necessity of the large dipole condition for \emph{both} black hole and microstate geometry.

\subsection{Geometries \texorpdfstring{$A$}{A} and \texorpdfstring{$B$}{B}}\label{sec:geometriesAB}
Here, we briefly repeat the four-center geometries called $A$ and $B$ from \cite{Bena:2020uup}. (The charges we give are related to the charges given in \cite{Bena:2020uup} by a gauge transformation, since we keep our moduli fixed as in (\ref{eq:MGharmfuncs}).) These are all exactly axisymmetric and (approximately) scaling; a parameter $\hat{k}$ can be dialed to an appropriate value to make the geometry approach the scaling point. Note that all centers are always smooth and so satisfy (\ref{eq:MGsmoothness}).

As discussed in \cite{Bena:2020uup} (especially Section 5.2 therein), for these geometries, the multipole ratios match very well, and the entropy parameter is small. We also give a few additional quantities that show that these geometries are indeed both of the large dipole variety.

\paragraph{Geometry \texorpdfstring{$A$}{A}}
This four-center scaling solution was first constructed in \cite{Heidmann:2017cxt}. The center charges are:
\be\begin{aligned}
 v_i &= \left( 1, 1, 12, -13 \right),\\
  k_i^1 &= \left( \frac{392157901841147 }{399439035817836}\hat{k}+\frac{118361894691555090254011}{17309024885439560000},\right.\\
  & \frac{392157901841147 }{399439035817836}\hat{k}+\frac{108975859584614420849511}{17309024885439560000},\\
  &\frac{425444488159300 }{33286586318153}\hat{k}+\frac{18958590398565735900719}{216362811067994500},\\
  &\left.-\frac{5098052723934911 }{399439035817836}\hat{k}-\frac{116071855576624493604643}{1331463452726120000} \right),\\
  k_i^2 &= \left(-\frac{20333393}{1250} ,-\frac{31240309}{2500},-\frac{945684581}{5000},\frac{1089498771}{5000}  \right),\\
  k_i^3 &= \left( \frac{251}{625},-\frac{11481}{2500},-\frac{287091}{5000},\frac{61609}{1000}  \right).
\end{aligned}\ee
The scaling solution is at:
\be \hat{k} \approx -0.804597.\ee
We give the center positions (chosen such that $M_1=0$) near the scaling point; here for $\hat{k} = -0.804$:
\be z_1 = 3.01459\times 10^{-9}, \qquad z_2 = 0.74\, z_1, \qquad z_3 = 0.87648\, z_1, \qquad z_4 = 0.944\, z_1.\ee

The corresponding supersymmetric black hole is clearly of the large dipole type:
\be \mathcal{H} = 7.74 \times 10^{-4}, \qquad \frac{D}{M_\text{BH}} = 35.8617,\ee
and we also give the angular momentum for the black hole for reference:
\be \label{eq:JaDgeomA} \frac{J_\text{BH}}{aD} = 4.757.\ee

The microstate geometry itself is clearly also of the large dipole variety since all centers are approximately at the same position $z_i=\mathcal{O}(1)\, z_1$. Explicitly, we can take $z_i=z_0+\tilde z_i$ with $z_0 = 2.68\times 10^{-9}$,  $\sum\tilde z_i = 0$ and $\tilde z_i\sim \mathcal{O}(10^{-10})$.
Further, the relevant microstate geometry angular momentum is given by:
\be -\frac{J_\text{MG}}{M_\text{BH} z_0} = 5.3741,\ee
which is indeed very close to (\ref{eq:JaDgeomA}).

\paragraph{Geometry \texorpdfstring{$B$}{B}}
The charges for this solution are given by:
\be \begin{aligned}
 v_i &= \left(1.000, -156.96, 159.0, -2.04 \right),\\
   k_i^1 &= \left(-58.32 + 1.94\, \hat{k}, 9014.42 - 147.19 \, \hat{k}, -9184.82 + 
 149.1\, \hat{k}, 113.0 - 1.91\, \hat{k}\right),\\
   k_i^2 &= \left(19.3019,-3361.53,3386.63,-44.4078\right),\\
   k_i^3 &= \left(18.4982,-2779.93,2797.31,-35.8799\right).
\end{aligned}\ee
The scaling solution is at:
\be \hat{k} \approx 0.5354.\ee
We give the center positions at $\hat{k} = 0.53$:
\be z_1 = -0.000325, \qquad z_2 = 1.00965\, z_1, \qquad z_3 = 1.000536\, z_1, \qquad z_4 = 0.995444\, z_1.\ee

The corresponding supersymmetric black hole is again clearly of the large dipole type:
\be \mathcal{H} = 7.9346\times 10^{-6}, \qquad \frac{D}{M_\text{BH}} = -251.1126,\ee
and we also give the angular momentum for the black hole for reference:
\be \label{eq:JaDgeomB} \frac{J_\text{BH}}{aD} = -11.644.\ee

The microstate geometry itself is clearly also of the large dipole variety; we can take $z_i=z_0+\tilde z_i$ with $z_0 = -3.25\times 10^{-4}$,  $\sum\tilde z_i = 0$ and $\tilde z_i\sim \mathcal{O}(10^{-6})$.
Finally, we have:
\be -\frac{J_\text{MG}}{M_\text{BH} z_0} = -11.656,\ee
which matches (\ref{eq:JaDgeomB}) very well.

Note that geometry $B$ and its corresponding black hole both have an order of magnitude larger dipoles than geometry $A$ and its corresponding black hole --- indeed, for $B$ we find $D/M\sim \mathcal{O}(100)$ and $z_0/\tilde z_i\sim \mathcal{O}(10^{-2})$ whereas for geometry $A$ we found $D/M\sim \mathcal{O}(10)$ and $z_0/\tilde z_i\sim\mathcal{O}(10^{-1})$. The matching of \direct and \indirect multipole ratios was also indeed much better for geometry $B$ compared to geometry $A$, as was discussed at length in \cite{Bena:2020uup}.

\section{Discussion and Further Conjectures}\label{sec:moreconj}

  Section \ref{sec:forward} shows clearly that having the property of ``large dipole'' (for both black hole and microstate geometry) implies that the multipole ratios calculated in both \direct and \indirect methods will match (in the precise sense of (\ref{eq:conjForward})). By contrast, the arguments presented in Section \ref{sec:3centres} are not completely conclusive. We have only considered a particular type of a three-centered geometry, whereas a generic proof would be for any number of centers with a generic profile --- and not just one where the centers behave as in (\ref{eq:2centreApprox}).
  
  However, the arguments used in Section \ref{sec:3centres} can in principle be generalized to e.g. a higher number of centers, although the details become much more complicated. Further, in all examples we have found or seen with $n\geq 3$ (of which e.g. geometries $A$ and $B$ of Section \ref{sec:geometriesAB} are prime examples), these arguments indeed hold. Thus, we have presented strong evidence for the new, more precise conjecture that the multipole ratios will match if and only if the microstate geometry \emph{and} black hole are of the large dipole type, in the sense of (\ref{eq:conjTotal}):
  \be \begin{aligned} \label{eq:conjTotalrepeat} \left(\frac{|D|}{M_\text{BH}}\gg 1\right) \text{ and } \left(\frac{z_0}{|\tilde z_i|}\gg 1\right) \text{ and } \left(\frac{J_\text{BH}}{aD} = -\frac{J_{\text{MG}}}{M_\text{BH}z_0}\right)\\
 \quad \Leftrightarrow \text{direct/indirect multipole ratios match}.
 \end{aligned}\ee
 It would be interesting to find a more physical interpretation of this result that can be linked to other properties of a microstate geometry --- to understand, for example, if having a ``large dipole'' is a ``typical'' property of a multicentered geometry. One possible hint is that our ``large dipole'' property is reminiscent of the ``biasing'' discussed in \cite{Bena:2006is} of pairs of centers that is needed to get large $J_L$ (in five dimensions).

  \medskip
  
In our analysis, based on all the examples we have found, we have also seen hints of how even a stronger statement may be true than the conjecture we have shown. We discuss here three such hints and possible stronger, refined conjectures that we strongly suspect are true but have not been able to prove (or find a counterexample). Future work can be the judge of whether these hints are the forebearers of new, interesting physics or mere red herrings.
\begin{itemize}
\item It seems as if the angular momentum condition (\ref{eq:Jsmatching}) is superfluous, i.e. that $|D|/M\gg 1$ and $z_0/\tilde z_i\gg 1$ themselves already \emph{imply} that (\ref{eq:Jsmatching}) holds, i.e.:
\be \label{eq:Jsmatchingrepeat} \frac{J_\text{BH}}{aD} \approx -\frac{J_{\text{MG}}}{M_\text{BH}z_0}.\ee
This is certainly the case in all of the large dipole geometries presented in this paper; we have not been able to construct any large dipole geometry where (\ref{eq:Jsmatchingrepeat}) does \emph{not} hold.

It is not clear exactly why (\ref{eq:Jsmatchingrepeat}) would hold for an arbitrary large dipole geometry --- the black hole quantity $J_\text{BH}/(aD)$ may be entirely fixed by the black hole charges --- see (\ref{eq:JoveraD}) --- but the same is not necessarily true for the microstate geometry $J_{\text{MG}}/(M_\text{BH}z_0)$, especially when the number of centers grows and thus the ways that $J_{\text{MG}}$ can be changed while keeping the overall charges $Q_I,P^I$ fixed increase. However, if true, this suggests that (\ref{eq:conjTotalrepeat}) could be strengthened to:
  \be \label{eq:conjstronger1} \left(\frac{|D|}{M_\text{BH}}\gg 1\right)  \text{ and } \left(\frac{z_0}{|\tilde z_i|}\gg 1\right)
  \ \Leftrightarrow\  \text{direct/indirect multipole ratios match}.
 \ee

\item It is clear that microstate geometries exist that are \emph{not} of the large dipole type for which the corresponding black hole \emph{is} of the large dipole type --- the geometry in Section \ref{sec:excounter} is an explicit example of such a geometry. However, we have not been able to find the converse, namely a large dipole microstate geometry which corresponds to a \emph{small} dipole black hole --- in all examples we have found, it seems that the microstate geometry having a large dipole automatically implies the large dipole for the corresponding black hole. If true, we could strengthen (\ref{eq:conjstronger1}) even further:
  \be  \label{eq:conjstronger2}\left(\frac{z_0}{|\tilde z_i|}\gg 1\right)
 \ \ \Leftrightarrow\ \ \text{direct/indirect multipole ratios match}.
 \ee
 This would be a remarkable and satisfying result. 

\item A final observation is that all microstate geometries that we have found that are of the large dipole type, so $z_0/\tilde z_i\gg 1$, seem to behave as an ``effective two-center geometry''. In essence, this means that there is a hierarchy of scales such that all centers are clustered around two centers. One can then  ``combine'' the centers into an ``effective'' two-center geometry. A multicentered geometry behaving as an effective two-center geometry seems like a fine-tuned situation, nevertheless e.g. geometries $A$ and $B$ of Section \ref{sec:geometriesAB}, as well as the three-center geometry of Section \ref{sec:limitMG}, all fall in this category --- without having been fine-tuned in any way. In fact, once again, we have never seen a microstate geometry that was of the large dipole type that did not behave effectively as a two-center geometry. Perhaps there is a link to be drawn with the observation in Section \ref{sec:BHtwocenter}, that the (non-extremal, rotating) black hole can also be seen as an effective two-center geometry with complex center positions. For example, one could imagine that the Euclidean Wick-rotated large dipole black hole and large dipole microstate geometries could be smoothly connected (in some sense) in the space of complex (but allowed \cite{Witten:2021nzp}) Euclidean metrics.
We leave investigating such connections to future work.

\end{itemize}




\section*{Acknowledgments}
 We would like to thank I. Bena for many discussions and encouragement during this work.
 DRM is supported by FWO Research Project G.0926.17N. This work is also partially supported by the KU Leuven C1 grant ZKD1118 C16/16/005.
 BG is supported in part by the ERC Grant 787320 - QBH Structure.

\appendix

\section{More Black Hole and Microstate Details}
In this appendix, we give additional details on the black holes and microstate geometries discussed in the main text.

\subsection{General Black Hole Charge Parameters}\label{app:chargeparams}
The general Chow-Comp\`ere \cite{Chow:2014cca} black hole discussed in Section \ref{sec:genBH} is determined by the 10 parameters $m,a,\delta_I,\gamma_I$ (for $I=0,1,2,3$). It is convenient to introduce the quantities $\mu_{1,2},\nu_{1,2}$ that are functions of the charge parameters $\delta_I,\gamma_I$ (see below for their explicit expressions).
Then, defining:
\be M_n = m\mu_1 + n \mu_2, \qquad N = m\nu_1 + n \nu_2,\ee
the physical parameters of the black hole can be written as:
\be \begin{aligned}
 M_\text{BH} &= M_n\left(n=-m\frac{\nu_1}{\nu_2}\right) = m\left(\mu_1 -\frac{\nu_1}{\nu_2}\mu_2\right),\\
 D &= m\left( \mu_2 +\frac{\nu_1}{\nu_2}\mu_1\right),\\
 J_\text{BH} &= m a \left(\frac{\nu_1^2}{\nu_2}+\nu_2\right),\end{aligned} \ee
where we have set $N=0$, giving us an asymptotically flat solution (otherwise $N$, the NUT charge, is an additional 11th parameter). The electromagnetic charges are given by:
\be Q_I = 2 \left(\frac{\partial M}{\partial \delta_I}\right)_{n=-m\nu_1/\nu_2}, \qquad P^I = -2\left(\frac{\partial N}{\partial \delta_I}\right)_{n=-m\nu_1/\nu_2}.\ee

Finally, we give the expressions for $\mu_{1,2},\nu_{1,2}$ in terms of $\delta_I,\gamma_I$. The following equations are all taken from (4.18), (4.19), (4.20), and (5.5) in \cite{Chow:2014cca} (with the range of $I$ changed here to $I=0,\cdots, 3$).
First, we define the shorthands:
\begin{align}
 s_{\delta I} &\equiv \sinh \delta_I, & c_{\delta I} &\equiv \cosh \delta_I,
 \end{align}
 and similarly for $s_{\gamma I},c_{\gamma I}$. We also define shorthands for products of these parameters such as:
 \be s_{\delta IJ} \equiv s_{\delta I} s_{\delta J},\ee
 and similar for products of $c_{\delta I},s_{\gamma I},s_{\gamma J}$; we can also specify more than two indices to multiply in such a product, such as $s_{\delta 0123}$.
 Then, the charge parameters $\mu_{1,2},\nu_{1,2}$ are given by:
\begin{align}
\mu_1 & = 1 + \sum_I \bigg( \frac{s_{\delta I}^2 + s_{\gamma I}^2}{2} - s_{\delta I}^2 s_{\gamma I}^2 \bigg) + \frac{1}{2} \sum_{I, J} s_{\delta I}^2 s_{\gamma J}^2 ,\\
\mu_2 & = \sum_I s_{\delta I} c_{\delta I} \bigg( \frac{s_{\gamma I}}{c_{\gamma I}} c_{\gamma 0 1 2 3} - \frac{c_{\gamma I}}{s_{\gamma I}} s_{\gamma 0 1 2 3}\bigg) ,\\
\nu_1 & = \sum_I s_{\gamma I} c_{\gamma I} \bigg( \frac{c_{\delta I}}{s_{\delta I}} s_{\delta 0 1 2 3} - \frac{s_{\delta I}}{c_{\delta I}} c_{\delta 0 1 2 3} \bigg) ,\\
\nu_2 &= \iota - \mathcal{D}
\end{align}
where
\begin{align}
\iota &= c_{\delta 0 1 2 3}c_{\gamma 0 1 2 3}+s_{\delta 0 1 2 3} s_{\gamma 0 1 2 3}+ \sum_{I < J} c_{\delta 0 1 2 3} \frac{s_{\delta I J}}{c_{\delta I J}} \frac{c_{\gamma I J}}{s_{\gamma I J}} s_{\gamma 0 1 2 3} , \\
\mathcal{D} &= c_{\delta 0 1 2 3}s_{\gamma 0 1 2 3}+s_{\delta 0 1 2 3}c_{\gamma 0 1 2 3} + \sum_{I < J} c_{\delta 0 1 2 3} \frac{s_{\delta I J}}{c_{\delta I J}} \frac{s_{\gamma I J}}{c_{\gamma I J}} c_{\gamma 0 1 2 3}.
\end{align}

The supersymmetric extremal limit in the static case, $a\rightarrow0$, is obtained by taking $\epsilon\rightarrow0$, while scaling the parameters as follows:
\begin{equation}
m\sim\epsilon^2,\quad \delta_I\sim\epsilon^0,\quad e^{\gamma_I}\sim\epsilon^{-1}.
\end{equation}

\subsection{Four-Dimensional Multicentered Geometries}\label{app:MGs}
The metric of the four-dimensional multicentered microstate geometry discussed in Section \ref{sec:MGs} is completely determined by the eight harmonic functions $(V,K^I, L_I, M)$:
\be ds^2 = - \mathcal{Q}^{-1/2}(dt +  \omega)^2 + \mathcal{Q}^{1/2}\left( dr^2 + r^2d\theta^2 + r^2\sin^2\theta d\phi^2\right),\ee
with the quartic invariant given by:
\begin{align} \label{eq:quarticinvdef} \mathcal{Q} &= Z_1 Z_2 Z_3 V - \mu^2 V^2,\\
 Z_I &= L_I + \frac12 C_{IJK} \frac{K^J K^K}{V}, \\ \mu &= M + \frac{1}{2V} K^I L_I + \frac{1}{6V^2}C_{IJK}K^I K^J K^K.
 \end{align}
 We work with $C_{IJK} = |\epsilon_{IJK}|$. The rotation one-form is determined by the differential equation on the 3D spatial base:
\be \label{eq:omegadiff} *_3 d\omega = V dM -M dV + \frac12\left( K^I dL_I - L_I dK^I\right),\ee
where $*_3$ is the three-dimensional Hodge star on the flat $\mathbb{R}^3$ basis. Together with the asymptotic condition that $\lim_{r\rightarrow\infty}\omega =0$, this differential equation completely determines $\omega$ (up to possible gauge transformations).

The \emph{bubble equations} are non-linear relations between the positions of the centers, the center charges, and the moduli; there is one for each center:
\be \label{eq:bubbleeqs} \sum_{j\neq i} \frac{\langle\Gamma^i, \Gamma^j\rangle}{ |\vec{r}_i - \vec{r}_j|} = \langle h, \Gamma^i\rangle\, ,
\ee
We have used the symplectic product of two charge vectors:
\be \label{eq:symplprod} \langle \Gamma^i, \Gamma^j \rangle \equiv m^i v^j - \frac12 k_I^i l_I^j - (i\leftrightarrow j).\ee
The moduli we use are always:
\be h = (1,0,0,0,1,1,1,0),\ee
which is sufficient to ensure $\mathcal{Q}\rightarrow 1$ so that the metric is indeed asymptotically flat.

Finally, for a given metric, one must additionally always explicitly check that the metric is everywhere regular, i.e.:
\be \mathcal{Q} \geq 0 .\ee

\section{Additional Useful Formulae}\label{app:multAngles}\label{app:CfuncsMG}

A trigonometric identity we use is:
\begin{align}
\sin(n\theta )&=\sum _{k{\text{ odd}}}(-1)^{\frac {k-1}{2}}{n \choose k}\cos ^{n-k}\theta \sin ^{k}\theta\notag\\
&=\sin \theta \sum _{i=0}^{(n+1)/2}\sum _{j=0}^{i}(-1)^{i-j}{n \choose 2i+1}{i \choose j}\cos ^{n-2(i-j)-1}\theta ,\\\cos(n\theta )&=\sum _{k{\text{ even}}}(-1)^{\frac {k}{2}}{n \choose k}\cos ^{n-k}\theta \sin ^{k}\theta =\sum _{i=0}^{n/2}\sum _{j=0}^{i}(-1)^{i-j}{n \choose 2i}{i \choose j}\cos ^{n-2(i-j)}\theta \,,
\end{align}

We also introduced a $C$-function for microstate geometries, given by:
\begin{multline}
C^{(1)}_{\ell m}(\varphi)=\frac{3^{-\frac{m}{2}}}{12\,(1+\varphi^2)\,\Gamma(1+m)}\sqrt{\frac{(\ell+m)!}{(\ell-m)!}}\Bigg(\frac{\varphi(1+\varphi^2)(3\,\varphi^2+1)}{(1-3\,\varphi^2)(1-\varphi^2)^2(3+\varphi^4)}\Bigg)^m\times\\
\big[2^{-m}(\varphi^4-4\,\varphi^2+7)(15\,\varphi^4-2\,\varphi^2+3)-(-1)^m(1+\varphi^2)^{2\,m}\big(3(3\,\ell+2)\varphi^4+2(6\,\ell-7)\varphi^2+3\,\ell\big)\big],\\
C^{(2)}_{\ell m}(\varphi)=\frac{3^{-\frac{m}{2}}}{12\,(1+\varphi^2)\,\Gamma(1+m)}\sqrt{\frac{(\ell+m)!}{(\ell-m)!}}\Bigg(\frac{\varphi(1+\varphi^2)(3\,\varphi^2+1)}{(1-3\,\varphi^2)(1-\varphi^2)^2(3+\varphi^4)}\Bigg)^m\times\\
\bigg[2^{-m}(\varphi^4-4\,\varphi^2+7)(-9\,\varphi^4+4\,\varphi^2-7)+(-1)^m(1+\varphi^2)^{2\,m}\bigg(7-4\,\varphi^2+9\,\varphi^4+\\
+\frac{3\,\ell(1+\varphi^2)(1+3\,\varphi^2)(3+\varphi^4)}{(1-\varphi^2)^2}\bigg)\bigg].
\end{multline}

\bibliographystyle{toine}
\bibliography{SUSY_multipole_ratios}

\end{document}